\documentclass[a4paper,11pt]{article}
\pdfoutput=1 

\usepackage{jheppub} 

\usepackage[T1]{fontenc} 
\usepackage{bm}
\usepackage{url}
\DeclareMathAlphabet{\pazocal}{OMS}{zplm}{m}{n}
\usepackage{braket}
\usepackage{amsmath}
\usepackage{amssymb}
\usepackage{leftidx}
\usepackage[colorinlistoftodos]{todonotes}
\usepackage{bm}
\usepackage{slashed}
\usepackage{calrsfs}
\usepackage{cancel}
\usepackage{dirtytalk}
\usepackage{upgreek}
\usepackage{hyperref}
\title{\boldmath (Non-)unitarity of strictly and partially massless fermions on de Sitter space}

\author[a]{Vasileios A. Letsios}

\affiliation[a]{Department of Mathematics, University of York\\Heslington, York, YO10 5DD, United Kingdom }

\emailAdd{vasileios.letsios@york.ac.uk}

\abstract{We present the dictionary between the one-particle Hilbert spaces of totally symmetric tensor-spinor fields of spin $s={3}/{2}, {5}/{2}$ with any mass parameter on $D$-dimensional ($D \geq 3$) de Sitter space ($dS_{D}$) and Unitary Irreducible Representations (UIR's) of the de Sitter algebra spin$(D,1)$. Our approach is based on expressing the eigenmodes on global $dS_{D}$ in terms of eigenmodes of the Dirac operator on the ${(D-1)}$-sphere, which provides a natural way to identify the corresponding representations with known UIR's under the decomposition spin$(D,1)$ $\supset$ spin$(D)$. Remarkably, we find that four-dimensional de Sitter space plays a distinguished role in the case of the gauge-invariant theories. In particular, the strictly massless spin-3/2 field, as well as the strictly and partially massless spin-5/2 fields on $dS_{D}$, are not unitary unless $D=4$. }

\begin{document} 
\maketitle
\flushbottom

\section{Introduction} \label{Introduction}
\subsection{Strictly and partially massless field theories in de Sitter space}
The de Sitter spacetime, apart from its relevance to inflationary cosmology, is also thought to be a good model for the asymptotic future of our Universe, as suggested by current experimental evidence in favor of a positive cosmological constant 
\cite{Perlmutter_1999,SloanDigitalSky,PlanckCollab}. The $D$-dimensional de~Sitter spacetime ($dS_{D}$) is the maximally symmetric solution of the vacuum Einstein field equations with positive cosmological constant $\Lambda$~\cite{hawking_ellis_1973}
\begin{equation}
    R_{\mu \nu}-\frac{1}{2} g_{\mu \nu} R + \Lambda  g_{\mu \nu}=0,
\end{equation}
where $g_{\mu \nu}$ is the metric tensor, $R_{\mu \nu}$ is the Ricci tensor and $R$ is the Ricci scalar. Throughout this paper we use units in which the cosmological constant is
\begin{equation}
    \Lambda=\frac{(D-2)(D-1)}{2 \hspace{1mm}},
\end{equation}
i.e. the de Sitter radius is one.

Unlike Minkowskian field theories, possible field theories of spin $s$ on $dS_{D}$ are not restricted to the two usual cases of massive and strictly massless theories, where for $D=4$ the former has $2s+1$ propagating degrees of freedom (DoF), while the latter has only 2 helicity DoF ($\pm s$) due to the gauge invariance of the theory~\cite{Tung}. On $dS_{D}$ there also exist intermediate gauge-invariant theories for $s\geq 2$, known as \textbf{partially massless}~\footnote{Partially massless theories exist also in anti-de Sitter spacetime. Partially and strictly massless theories on both de Sitter and anti-de Sitter spacetimes are discussed in Ref.~\cite{Deser_Waldron_phases}.} theories~\cite{Deser_Waldron_null_propagation, Deser_Waldron_stability_of_massive_cosm, Deser_Waldron_phases, Deser_Waldron_partial_masslessness, Deser_Waldron_Conformal}. For a given spin $s \geq 1$, there exists one strictly massless theory and $[s]-1$ different partially massless theories, where $[s]=s$ if the spin $s$ is an integer and $[s]=s-1/2$ if $s$ is a half-odd integer. Partial masslessness was first observed for the spin-2 field by Deser and Nepomechie~\cite{DESER_NEPOM_1,DESER_NEPOM_2} and for higher integer-spin fields by Higuchi~\cite{STSHS}. Partially massless theories with various spins have been discussed further in a series of papers by Deser and Waldron~\cite{ Deser_Waldron_null_propagation, Deser_Waldron_stability_of_massive_cosm, Deser_Waldron_phases, Deser_Waldron_partial_masslessness, Deser_Waldron_Conformal, Deser_Waldron_ArbitrarySR}. Note that this paragraph, as well as the rest of the paper, refers only to totally symmetric tensor and tensor-spinor fields. Mixed-symmetry tensor fields on $dS_{D}$ - for which strict and partial masslessness also occur - have been discussed in Ref.~\cite{Mixed_Symmetry_dS}.

Each strictly or partially massless theory of spin $s$ is conveniently labeled by a distinct value of the `depth' $\uptau=1,2,...,[s]$ (where the value $\uptau=1$ corresponds to strict masslessness) and in 4 dimensions there are $2\uptau$ propagating helicities, namely: $(\pm s, \pm (s-1), ..., \pm (s-\uptau+1))$~\cite{Deser_Waldron_phases, Deser_Waldron_null_propagation, Deser_Waldron_partial_masslessness}. For given spin $s$ and depth $\uptau$, each of these gauge-invariant theories corresponds to a distinct tuning of the mass parameter to the cosmological constant $\Lambda$~\cite{STSHS,Deser_Waldron_phases,Deser_Waldron_null_propagation,Deser_Waldron_ArbitrarySR,Deser_Waldron_partial_masslessness}. Higuchi classified the tunings of the mass parameter for all strictly and partially massless theories with arbitrary integer spin by studying the group-theoretic properties of the eigenmodes of the Laplace-Beltrami operator on $dS_{D}$~\cite{STSHS,Yale_Thesis}. Deser and Waldron gave an analogous classification for arbitrary integer and half-odd-integer spins by using group representation methods based on the de Sitter/CFT correspondence~\cite{Deser_Waldron_ArbitrarySR}.

\subsection{Eigenmodes, `field theory-representation theory' dictionary and purpose of this paper}
Unitarity of field theories is very important for physical problems since it ensures the positivity of probabilities. 
A sufficient condition for field-theoretic unitarity on $dS_{D}$ is that of the unitarity of the underlying representation of the de Sitter (dS) algebra, spin$(D,1)$. Particles in a $D$-dimensional dS universe correspond to Unitary Irreducible Representations (UIR's) of spin$(D,1)$.

\noindent \textbf{Representation-theoretic insight from eigenmodes.} The interplay between free field theory on $dS_{D}$ and representation theory of spin$(D,1)$ manifests beautifully itself in the solution space - consisting of eigenmodes - of the corresponding field equation.\footnote{If a dS invariant positive-definite scalar product exists for the eigenmodes, then the vector space of eigenmodes can be identified with the one-particle Hilbert of the corresponding unitary quantum field theory.} {Let us briefly discuss Higuchi's work~\cite{Yale_Thesis, STSHS} in order to demonstrate the great amount of representation-theoretic knowledge that we can obtain for a free field theory on $dS_{D}$ by studying its eigenmodes. In particular, in Refs.~\cite{Yale_Thesis, STSHS} Higuchi studied the group-theoretic properties of totally symmetric tensor eigenmodes of the Laplace-Beltrami operator on $dS_{D}$ ($D \geq 3$). In these works, he showed that the phenomenon of partial masslessness exists for all totally symmetric tensor fields of spin $s \geq 2$ on $dS_{D}$ by detecting pure gauge modes (these eigenmodes indicate the gauge invariance of the theory). Also, by calculating the norm of the physical strictly/partially massless eigenmodes using a dS invariant scalar product, he showed that all strictly and partially massless theories with arbitrary integer spin $s$ are unitary for all $D \geq 3$. Moreover, he showed that for all integer spins there exist mass (parameter) ranges where the eigenmodes have negative norm - i.e. the corresponding spin$(D,1)$ representations are non-unitary. The unitary strictly/partially massless theories appear at special tunings of the mass parameter corresponding to the boundaries of the `forbidden' mass ranges - see Deser and Waldron's works for a detailed analysis and a physical insight into these `forbidden' ranges~\cite{ Deser_Waldron_null_propagation, Deser_Waldron_stability_of_massive_cosm, Deser_Waldron_phases, Deser_Waldron_partial_masslessness}. Last, Higuchi's group-theoretic analysis of the eigenmodes showed that there is a lower bound for the mass parameter of integer-spin fields, below which the fields can only be non-unitary\footnote{The Higuchi bound depends on both the (integer) spin of the field and the spacetime dimension $D$~\cite{STSHS}.}. This bound is known as the `Higuchi bound' in the modern literature - see, e.g Ref.~\cite{Lust, Higuchiforb}.} 

\noindent \textbf{`Field theory-representation theory' dictionary and a gap in the literature.}
The basis elements of spin$(D,1)$ correspond to the $(D+1)D/2$ Killing vectors of $dS_{D}$ and they act on eigenmodes in terms of Lie derivatives (or spinorial generalizations thereof~\cite{Kosmann, Ortin}). The (spinorial) Lie derivatives with respect to Killing vectors commute with the field equation of the free theory~\cite{Kosmann, Ortin} and the solution space is identified with the representation space of a - often irreducible - representation of spin$(D,1)$~\cite{Yale_Thesis,STSHS}. What we would like to know is whether this representation, which is formed by eigenmodes, is unitary. Fortunately, all UIR's of spin$(D,1)$ have been classified by Ottoson and Schwarz~\cite{Ottoson, Schwarz}~(see also Refs.~\cite{Wong, Hirai1,Hirai2}). Thus, as field theorists, we would like to construct a dictionary between the known UIR's of spin$(D,1)$ and eigenmode spaces (i.e. one-particle Hilbert spaces) of free field theories on $dS_{D}$. Such a dictionary was first constructed by Higuchi~\cite{Yale_Thesis} for totally symmetric integer-spin fields\footnote{See also Refs.~\cite{Sun, Gizem} for more recent discussions concerning the `field theory-representation theory' dictionary for integer-spin fields on $dS_{D}$.} and was later extended to mixed-symmetry integer-spin fields by Basile, Bekaert and Boulanger~\cite{Mixed_Symmetry_dS}. However, a detailed study of the dictionary for tensor-spinor fields for arbitrary $D$ is absent from the literature\footnote{For $D=4$, a dictionary for half-odd-integer-spin fields has been obtained in Ref.~\cite{Gazeau}. }. 

\noindent \textbf{Main aim.} It is the purpose of the present article to construct the dictionary between one-particle Hilbert spaces (consisting of eigenmodes) and UIR's of spin$(D,1)$ for the vector-spinor (i.e. spin-3/2) field and symmetric rank-2 tensor-spinor (i.e. spin-5/2) field on $dS_{D}$.
\subsection{Main result for strictly and partially massless theories of spin \texorpdfstring{$s=3/2, 5/2$}{..}}
 The dictionary between one-particle Hilbert spaces of unitary spin-$s=3/2, 5/2$ field theories on $dS_{D}$ and UIR's of spin$(D,1)$ will be given in Section~\ref{Sec_Dictionary} (for both massive and strictly/partially massless fields). However, here we would like to draw attention to our remarkable main result concerning the strictly and partially massless theories:
\begin{itemize}
    \item \textbf{Main result:} The strictly massless spin-3/2 field (gravitino field) and the strictly and partially massless spin-5/2 fields on $dS_{D}$ ($D \geq 3$) are not unitary unless $D=4$.
\end{itemize}
(The case with $D=2$ is not discussed in the present article.) As we will see later, our analysis for the spin-3/2 and spin-5/2 cases suggests that our main result should hold for all strictly and partially massless fields with half-odd-integer spin $s \geq 3/2$.

According to our main result, four-dimensional dS space plays a distinguished role in the unitarity of the strictly massless spin-3/2 field and the strictly and partially massless spin-5/2 fields. This is an example of a remarkable and previously unknown feature of dS field theory that has no known field-theoretic counterparts in anti-de Sitter and Minkowski spacetimes. As will become clear, the significance of four-dimensional dS space is related to the representation theory of spin$(D,1)$, where the latter allows (totally symmetric) fermionic strictly/partially massless UIR's only for $D=4$ (corresponding to a direct sum of spin$(4,1)$ UIR's in the Discrete Series - see Section~\ref{Sec_Dictionary}). Also, although it might be a mere mathematical coincidence, it is interesting that the dimensionality that plays a special representation-theoretic role happens to correspond to the number of the observed macroscopic dimensions of our Universe. 

\subsection{Strategy}
Our strategy in order to construct the dictionary between spin$(D,1)$ UIR's and spin-$s=3/2, 5/2$ one-particle Hilbert spaces on $dS_{D}$ is based on constructing the dS eigenmodes using the method of separation of variables~\cite{Camporesi, CHH, Letsios}. More specifically, we are going to express the spin-3/2 and spin-5/2 eigenmodes on global $dS_{D}$ in terms of tensor-spinor eigenmodes of the Dirac operator on $S^{D-1}$. This will help us determine the spin$(D)$ content of the spin$(D,1)$ representations formed by the eigenmodes on $dS_{D}$ - by spin$(D)$ content we mean the irreducible representations of spin$(D)$ that appear in a spin$(D,1)$ representation under the decomposition spin$(D,1) \supset $ spin$(D)$~\cite{Ottoson, Schwarz}. We will also obtain the values of the spin$(D,1)$ quadratic Casimir corresponding to the eigenmodes on $dS_{D}$. Once we have determined both the quadratic Casimir and the spin$(D)$ content for the representations formed by the dS eigenmodes, we will be able to construct the dictionary between one-particle Hilbert spaces and UIR's of spin$(D,1)$ by using the known classification of UIR's~\cite{Ottoson, Schwarz} under the decomposition spin$(D,1) \supset $ spin$(D)$. We also provide the dictionary for the spin-1/2 field (as the group-theoretic properties of the spin-1/2 eigenmodes on global $dS_{D}$ have been already studied by the author~\cite{Letsios}), while our analysis also allows us to propose a dictionary for totally symmetric tensor-spinors of any spin $s \geq 3/2$. 

 As for our main result concerning the strictly/partially massless theories of spin $s=3/2,5/2$, we will show that for $D \neq 4$ there is a mismatch between the values of the quadratic Casimir for the strictly/partially massless eigenmodes and the values corresponding to the UIR's of spin$(D,1)$ and/or another mismatch between the representation labels of the eigenmodes and the allowed labels in spin$(D,1)$ UIR's. (The spin$(D,1)$ representation labels we use in this paper specify a spin$(D,1)$ representation under the decomposition spin$(D,1) \supset $ spin$(D)$~\cite{Ottoson, Schwarz, Yale_Thesis, STSHS} and their role is similar to the role played by the highest weights in spin$(D+1)$ representations - see Section~\ref{Sec_Classification_UIRs}.) In other words, we will demonstrate that there are no UIR's of spin$(D,1)$ that correspond to the strictly massless spin-3/2 field and to the strictly and partially massless spin-5/2 fields on $dS_{D}$ for $D \neq 4$. However, for $D=4$, both the quadratic Casimir and the representation labels of the strictly/partially massless theories correspond to the Discrete Series UIR's of spin$(4,1)$. 

\noindent \textbf{An alternative technical explanation.} A technical explanation of all the results reported in this paper can be given by studying the (non-)existence of positive-definite dS invariant scalar products for the spin-3/2 and spin-5/2 eigenmodes on $dS_{D}$. Such an analysis has been carried out in detail by the author and will be presented in a separate article~\cite{Letsios_in_progress, Letsios_arxiv_long}, in which the author has extended Higuchi's methods~\cite{Yale_Thesis,STSHS} to the case of spin-3/2 and spin-5/2 eigenmodes on $dS_{D}$ ($D \geq 3$). In particular, in Refs.~\cite{Letsios_in_progress, Letsios_arxiv_long} the author has proved the following results for the strictly/partially eigenmodes of spin $s=3/2,5/2$ on $dS_{D}$ ($D \geq 3$):
\begin{itemize}
    \item For odd $D$ all  dS invariant scalar products are identically zero. 
    \item For even $D > 4$ all dS invariant scalar products are indefinite giving always rise to positive-norm and negative-norm eigenmodes that mix with each other under spin$(D,1)$ boosts.
    \item The $D=4$ case is special as the positive-norm sector decouples from the negative-norm sector. Then, both sectors can be viewed as  positive-norm sectors and each sector independently forms a spin$(4,1)$ UIR in the Discrete Series.
\end{itemize}
 Although we have not performed such a technical analysis for the eigenmodes with half-odd-integer spin $s \geq 7/2$, the analysis of our present paper suggests that our main result extends to all strictly and partially massless fields with half-odd-integer spin $s \geq 7/2$ on $dS_{D}$.


\subsection{Outline of the paper, notation and conventions}
The rest of the paper is organised as follows. In Section~\ref{background_dS}, we begin by presenting the basics about tensor-spinor fields on $dS_{D}$ (gamma matrices, vielbein fields, spin connection, and the spinorial generalisation of the Lie derivative) and, then, we specialise to the global slicing of $dS_{D}$. In Section~\ref{Sec_Classification_UIRs}, we review the classification of the spin$(D,1)$ UIR's under the decomposition spin$(D,1)\supset$ spin$(D)$ given originally in Refs.~\cite{Ottoson, Schwarz}. In Section~\ref{Section_dS_eigenmodes}, we begin by discussing the totally symmetric tensor-spinor eigenmodes of the Dirac operator on $S^{D-1}$ that are also gamma-traceless and divergence-free, as well as the way they form representations of spin$(D)$~(Subsection~\ref{Sub_Sec_eigenmodes_spheres}). Then, using the aforementioned eigenmodes on $S^{D-1}$, we present the construction of the TT eigenmodes of the spin-3/2 field on $dS_{D}$ for both even $D \geq 4 $~(Subsecton~\ref{Sub_Sec_separate_spin3/2_even}) and odd $D \geq 3$~(Subsection~\ref{Sub_Sec_separate_spin3/2_odd}), in order to illustrate the method of separation of variables for tensor-spinor fields. The spin$(D)$ content of the spin$(D,1)$ representations formed by the spin-3/2 eigenmodes is also identified and the main results are tabulated in Tables 1 and 2. In Subsection~\ref{Sub_Sec_separate_spin5/2}, we present our basic results concerning the TT eigenmodes for the spin-5/2 field on $dS_{D}$ ($D \geq 3$). In Section~\ref{Sec_quadratic_Casimir}, we obtain the quadratic Casimir for the spin$(D,1)$ representation formed by eigenmodes with half-odd-integer spin $s \geq 1/2$ on $dS_{D}$ by using ``analytic continuation" techniques that relate $dS_{D}$ to $S^{D}$. In Section~\ref{Sec_main_result_unitarity}, after identifying the pure gauge and physical modes of our strictly/partially massless theories~(Subsection~\ref{Subsection_puregauge_phys}), we prove the main result of this paper, i.e. the strictly massless spin-3/2 field, as well as the strictly and partially massless spin-5/2 fields on $dS_{D}$, are not unitary unless $D=4$~(Subsection~\ref{Subsec_nonunitarity}). In order to achieve this, we take advantage of both the spin$(D)$ content and the quadratic Casimir corresponding to our physical modes on $dS_{D}$ and then we show that they do not agree with any UIR of spin$(D,1)$ unless $D=4$. In Section~\ref{Sec_Dictionary}, we present our dictionary between spin$(D,1)$ UIR's and (totally symmetric) tensor-spinor fields with arbitrary mass parameters on $dS_{D}$ ($D \geq 3$). Although in the main part of the present paper we discuss the spin-3/2 and spin-5/2 fields, our analysis allows us to propose a dictionary for all (totally symmetric tensor-)spinor fields with spin $s \geq 1/2$.

\textbf{Notation and conventions.} We use the term `tensor-spinor field of rank $r$' in order to refer to a $r^{th}$-rank tensor where each one of its tensor components is a spinor. Other authors prefer the name spinor-tensors for these objects - see, e.g., Ref.~\cite{CHH}. We use the mostly plus metric sign convention for $dS_{D}$. Lowercase Greek tensor indices refer to components with respect to the `coordinate basis' on $dS_{D}$. Coordinate basis tensor indices on $S^{D-1}$ are denoted as $\tilde{\mu}_{1}, \tilde{\mu}_{2},...\,$. Lowercase Latin tensor indices are `flattened', i.e. they refer to components with respect to the vielbein basis (the indices $a,b,c,d,f$ run from $0$ to $D-1$, while the indices $i,j,k$ run from $1$ to $D-1$). Summation over repeated indices is
understood. We denote the symmetrisation of a pair of indices as $A_{(\mu \nu)} \equiv  (A_{\mu \nu}+A_{\nu \mu})/2$ and the anti-symmetrisation as $A_{[\mu \nu]} \equiv  (A_{\mu \nu}-A_{\nu \mu})/2$.
 Spinor indices are always suppressed throughout this paper. The rank of tensor-spinors on $dS_{D}$ is denoted as $r$, while the rank of tensor-spinors on $S^{D-1}$ as $\tilde{r}$. The complex conjugate of the complex number $z$ is $z^{*}$.

\section{Background material concerning tensor-spinors on \texorpdfstring{$dS_{D}$}{dS}} \label{background_dS}
Fermionic fields with arbitrary half-odd-integer spin $s \equiv r+1/2$ and mass parameter $M$ on $dS_{D}$ can be described by totally symmetric tensor-spinors $\Psi_{\mu_{1}...\mu_{r}}$ satisfying the onshell conditions~\cite{Deser_Waldron_null_propagation,Deser_Waldron_ArbitrarySR}:
\begin{align}
   &\left( \slashed{\nabla}+M\right)\Psi_{\mu_{1}...\mu_{r}}=0  \label{Dirac_eqn_fermion_dS}\\
   & \nabla^{\alpha}\Psi_{\alpha \mu_{2}...\mu_{r}}=0, \hspace{4mm}  \gamma^{\alpha}\Psi_{\alpha \mu_{2}...\mu_{r}}=0, \label{TT_conditions_fermions_dS}
\end{align}
where $\slashed{\nabla}=\gamma^{\nu}\nabla_{\nu}$ is the Dirac operator. From now on, we will refer to the divergence-free and gamma-tracelessness conditions in eq.~(\ref{TT_conditions_fermions_dS}) as the TT conditions.

 The half-odd-integer-spin theories described by eqs.~(\ref{Dirac_eqn_fermion_dS}) and (\ref{TT_conditions_fermions_dS}) become gauge-invariant (i.e. strictly/partially massless) for the following imaginary values of the mass parameter $M= i \tilde{M}$~\cite{Deser_Waldron_ArbitrarySR}:
\begin{align}\label{values_mass_parameter_masslessness_fermion}
    \tilde{M}^{2}=-M^{2}=\left(r-\uptau+\frac{D-2}{2}\right)^{2} \hspace{6mm}(\uptau=1,...,r)
\end{align}
for $r \geq 1$ (i.e. $s \geq 3/2$).
Real values of $M$ - including $M=0$ - correspond to non-gauge-invariant theories.

\subsection{Gamma matrices, vielbein fields, spin connection and Lie-Lorentz derivative on \texorpdfstring{$dS_{D}$}{dSD}}
 The $2^{[D/2]}$-dimensional\footnote{For $D$ even we have $[D/2]=D/2$. For $D$ odd we have $[D/2]=(D-1)/2$.}~gamma matrices $\gamma^{a}$ (with `flattened' indices $a = 0,1,...,D-1$) satisfy the anti-commutation relations 
\begin{equation}\label{anticommutation_relations_gamma}
   \{\gamma^{a}, \gamma^{b}\}  = 2 \eta^{ab} \bm{1}, \hspace{10mm} a,b=0,1,...,D-1,
\end{equation}
where $\bm{1}$ is the spinorial identity matrix and $\eta^{a b}= diag(-1,1,...,1)$.
 The vielbein fields $\bm{e}_{a}=e^{\mu}{\hspace{0.2mm}}_{a}\partial_{\mu}$, determining an orthonormal frame, satisfy
\begin{align}
    e_{\mu}{\hspace{0.2mm}}^{a} \, e_{\nu}{\hspace{0.2mm}}^{b}\eta_{ab}=g_{\mu \nu}, \hspace{4mm}e^{\mu}{\hspace{0.2mm}}_{a}\,e_{\mu}{\hspace{0.2mm}}^{b}=\delta^{b}_{a},
\end{align}
where the co-vielbein fields $\bm{e}^{a}=e_{\mu}{\hspace{0.2mm}}^{a}\,dx^{\mu}$ define the dual coframe. The gamma matrices with coordinate basis indices are defined using the vielbein fields as $\gamma^{\mu}(x) \equiv e^{\mu}{\hspace{0.2mm}}_{a}(x) \gamma^{a}$.

The covariant derivative for a vector-spinor field is 
  \begin{equation}\label{covariant_deriv_vector_spinor}
      \nabla_{\nu} \Psi_{\mu} = \partial_{\nu}  \Psi_{\mu}  + \frac{1}{4} \omega_{\nu bc} \gamma^{bc}  \Psi_{\mu}-\Gamma^{\lambda}_{\hspace{1mm}\nu \mu} \Psi_{\lambda},
  \end{equation}
where $\omega_{\nu b c  }=\omega_{\nu [b c]  } =e_{\nu}{\hspace{0.2mm}}^{a}\omega_{a b c  }$ is the spin connection, $\Gamma^{\lambda}_{\hspace{1mm}\nu \mu}$ are the Christoffel symbols and $\gamma^{bc} = \gamma^{[b}  \gamma^{c]}$. The covariant derivatives for higher-spin tensor-spinors are given by straightforward generalisations of eq.~(\ref{covariant_deriv_vector_spinor}). It is easy to check that the gamma matrices are covariantly constant, as $\nabla_{\mu} \gamma^{a}=\tfrac{1}{4}\omega_{\mu b c} (\gamma^{bc} \gamma^{a} - \gamma^{a} \gamma^{bc} )+ \omega_{\mu}\hspace{0.1mm}^{a}\hspace{0.1mm}_{c} \gamma^{c}=0$. According to our sign convention, we have~\footnote{The sign convention we use for the spin connection is the opposite of the one used in Refs.~\cite{Camporesi, Letsios}.} 
 \begin{equation}
      \partial_{\mu} e^{\rho}\hspace{0.1mm}_{b} + {\Gamma}^{\rho}_{\mu \sigma}e^{\sigma}\hspace{0.1mm}_{b} - \omega_{\mu}\hspace{0.1mm}^{c}\hspace{0.1mm}_{b}  \,e^{\rho}\hspace{0.1mm}_{c}=0.
  \end{equation}

For each value of the mass parameter $M$ in eq.~(\ref{Dirac_eqn_fermion_dS}), the set of TT eigenmodes $\Psi_{\mu_{1}...\mu_{r}}$  forms a representation of the de Sitter algebra spin$(D,1)$, which - as we will see below - may be unitary or non-unitary depending on both $M$ and the dimension $D$. The Killing vectors generating spin$(D,1)$ act on tensor-spinors in terms of the spinorial generalisation of the Lie derivative~\cite{Kosmann, Ortin} - also known as Lie-Lorentz derivative - as:
\begin{align}\label{Lie_Lorentz}
 \mathbb{L}_{{\xi}}~{\Psi}_{\mu_{1}...\mu_{r}}  =~&  \xi^{\nu} \nabla_{\nu} {\Psi}_{ \mu_{1}...\mu_{r}} +{\Psi}_{ \nu \mu_{2}...\mu_{r}} \nabla_{\mu_{1}}\xi^{\nu}+{\Psi}_{ \mu_{1} \nu \mu_{3}...\mu_{r}} \nabla_{\mu_{2}}\xi^{\nu}+...+ {\Psi}_{ \mu_{1} ...\mu_{r-1}\nu}\nabla_{\mu_{r}}\xi^{\nu}\nonumber \\
 &+ \frac{1}{4}  \nabla_{\kappa} \xi_{\lambda}  \gamma^{\kappa} \gamma^{\lambda }   {\Psi}_{\mu_{1}...\mu_{r}},
\end{align}
where $\xi^{\mu}$ is any dS Killing vector - i.e. $\nabla_{(\mu}  \xi_{\nu)}=0$. The Lie-Lorentz derivative satisfies~\cite{Ortin}
\begin{subequations}
\begin{align}
&\mathbb{L}_{\xi}~e_{\mu}^{\hspace{2mm}a}=0 ,\\
& \mathbb{L}_{\xi}~\gamma^{a}=0,  
\end{align}
\end{subequations}
 as well as
 \begin{align}\label{commutator_cov_LieLorentz}
     \left( \mathbb{L}_{{\xi}} \nabla_{\mu}-\nabla_{\mu}\mathbb{L}_{\xi}\right)~\Psi_{\mu_{1}...\mu_{r}}=0,
 \end{align}
 and hence $\mathbb{L}_{{\xi}}$ commutes with the Dirac operator. Moreover, the Lie-Lorentz derivative preserves the Lie bracket between any two vectors $\xi^{\mu}, X^{\mu} \in$ spin$(D,1)$ as
 \begin{align}
       \left( \mathbb{L}_{{\xi}} \mathbb{L}_{{X}} -\mathbb{L}_{{X}}\mathbb{L}_{{\xi}} \right) \Psi_{\mu_{1}... \mu_{r}}=\mathbb{L}_{{[\xi,X]}} \Psi_{\mu_{1}... \mu_{r}}.
 \end{align}

As for the representation of our gamma matrices on $dS_{D}$, we choose the following:
\begin{itemize}
    \item \textbf{For $\bm{D}$ even}: the $2^{D/2}$-dimensional gamma matrices are
\begin{equation}\label{even_gammas}
 \gamma^{0}=i \begin{pmatrix}  
   0 & \bm{1} \\
   \bm{1} & 0
    \end{pmatrix} , \hspace{5mm}
    \gamma^{j}=\begin{pmatrix}  
   0 & i\widetilde{\gamma}^{j} \\
   -i\widetilde{ \gamma}^{j} & 0
    \end{pmatrix} ,
     \end{equation} 
($ j=1,..., D-1$) where the $2^{(D-2)/2}$-dimensional gamma matrices $\widetilde{\gamma}^{j}$ generate a Euclidean Clifford algebra in $D-1$ dimensions, as 
\begin{equation} \label{Euclidean_Clifford_relns}
    \{ \widetilde{\gamma}^{j}, \widetilde{\gamma}^{k}\} = 2 \delta^{jk} \bm{1}, \hspace{7mm}j,k=1,...,D-1.
\end{equation}
 One can construct the extra gamma matrix $\gamma^{D+1}$ which is given by the product $\gamma^{D+1} \equiv \epsilon \,\gamma^{1}\gamma^{2}...\gamma^{D-1}\gamma^{0}$, where $\epsilon$ is a phase factor. The matrix $\gamma^{D+1}$ anti-commutes with each of the $\gamma^{a}$'s in eq.~(\ref{even_gammas}). We choose the phase factor $\epsilon$ such that
\begin{align}\label{gamma(D+1)}
    \gamma^{D+1}=\begin{pmatrix}
    \bm{1} & 0\\
    0      & -\bm{1}
    \end{pmatrix}.
\end{align}
For $D=4$ this is the familiar matrix $\gamma^{5}$.
\item \textbf{For $\bm{D}$ odd}: the $2^{(D-1)/2}$-dimensional gamma matrices are
\begin{equation}\label{odd_gammas}
   \gamma^{0}=i \begin{pmatrix}  
    \bm{1} & 0 \\
 0& -\bm{1} 
    \end{pmatrix} ,\hspace{7mm}
    \gamma^{j}= \widetilde{\gamma}^{j},\hspace{8mm}j=1,...,D-1,
\end{equation}
where the $\widetilde{\gamma}^{j}$'s are $2^{(D-1)/2}$-dimensional gamma matrices generating a Euclidean Clifford algebra in $D-1$ dimensions (see eq.~(\ref{Euclidean_Clifford_relns})).
\end{itemize}


\subsection{Specialising to global coordinates}
In order to obtain explicit expressions for the TT eigenmodes of the field equation~(\ref{Dirac_eqn_fermion_dS}), we will choose to work with the global slicing of $dS_{D}$. In global coordinates the line element is 
\begin{equation} \label{dS_metric}
    ds^{2}=-dt^{2}+\cosh^{2}{t} \,d\Omega^{2}_{D-1},
\end{equation}
($t \in \mathbb{R}$) where $d\Omega^{2}_{D-1}$ is the line element of $S^{D-1}$. The line element of $S^{m}$ can be parameterised as  
\begin{align} \label{line_element_sphere_inductive}
    d\Omega^{2}_{m}=d\theta^{2}_{m} + \sin^{2}{\theta_{m}} d\Omega_{m-1}^{2},\hspace{8mm} m=2,3,...,D-1,
\end{align} 
 with $0 \leq \theta_{m} \leq  \pi$, while $d\Omega_{m-1}^{2}$ is the line element of $S^{m-1}$. For $m=1$ we have $d\Omega_{1}^{2}=d \theta_{1}^{2}$ with $0 \leq \theta_{1} \leq 2 \pi$. We will use the symbol $\bm{\theta_{D-1}} \equiv (\theta_{D-1}, \theta_{D-2},...,\theta_{1})$ to denote a point on $S^{D-1}$.
 
The non-zero Christoffel symbols on global $dS_{D}$ are 
\begin{align}\label{Christoffels_dS}
    &\Gamma^{t}_{\hspace{0.2mm}\theta_{i} \theta_{j}}=\cosh{t} \sinh{t} \hspace{1mm}\tilde{g}_{\theta_{i} \theta_{j}}, \hspace{2mm} \Gamma^{\theta_{i}}_{\hspace{0.2mm}\theta_{j} t} =\tanh{t}  \hspace{1mm}\tilde{g}^{\theta_{i}}_{\theta_{j}}, \nonumber \\ 
& \Gamma^{\theta_{k}}_{\hspace{0.2mm}\theta_{i} \theta_{j}}=\tilde{\Gamma}^{\theta_{k}}_{\hspace{0.2mm}\theta_{i} \theta_{j}},
\end{align}
where $\tilde{g}_{\theta_{i} \theta_{j} }$ and $\tilde{\Gamma}^{\theta_{k}}_{\hspace{0.2mm}\theta_{i} \theta_{j}}$ are the metric tensor and the Christoffel symbols, respectively, on $S^{D-1}$. We choose the following expressions for the vielbein fields on $dS_{D}$:
\begin{equation}\label{vielbeins}
    e^{t}{\hspace{0.2mm}}_{0}=1, \hspace{5mm}  e^{\theta_{i}}{\hspace{0.2mm}}_{i}=\frac{1}{\cosh{t}} \tilde{e}^{\theta_{i}}{\hspace{0.2mm}}_{i} , \hspace{7mm}i=1,...,D-1,
\end{equation}
 where $\tilde{e}^{\theta_{i}}{\hspace{0.2mm}}_{i}$ are the vielbein fields on $S^{D-1}$. The non-zero components of the spin connection on $dS_{D}$ are given by
\begin{equation}\label{spin_connection_dS}
   \omega_{ijk} = \frac{  \tilde{\omega}_{ijk}}{ \cosh{t}  } , \hspace{6mm}  \omega_{i0k}= -\omega_{ik0}=  -\tanh{t}\hspace{1mm} \delta_{ik}, \hspace{9mm} i,j,k=1,...,D-1,
        \end{equation} 
where $\tilde{\omega}_{ijk}$ are the spin connection components on $S^{D-1}$.





\section{Classification of the UIR's of \texorpdfstring{spin$(D,1)$}{spin(D,1)}} \label{Sec_Classification_UIRs}
Here we review the classification of the spin$(D,1)$ UIR's by Ottoson~\cite{Ottoson} and Schwarz~\cite{Schwarz}. 
These authors have classified the UIR's of spin$(D,1)$ under the decomposition spin$(D,1) \supset$ spin$(D)$ - in the present paper spin$(D)$ denotes the Lie algebra of SO$(D)$. Under this decomposition, an irreducible representation of spin$(D)$ appears at most once in a UIR of spin$(D,1)$~\cite{Dixmier}. The case with $D=2p$ and the case with $D=2p+1$, where $p$ is a positive integer, are studied separately. Below we will adopt the notation for the labels of UIR's that were used by Higuchi in Ref.~\cite{Yale_Thesis}. However, we will use the names of the UIR's that are used in the modern literature~\cite{Mixed_Symmetry_dS, Sun, Gizem}.

\noindent \textbf{Representations of spin$\bm{(D)}$.}
Let us review the basics concerning spin$(D)$ representations. As is well-known, a representation of spin$(2p)$ or spin$(2p+1)$ is specified by the highest weight of the representation~\cite{barut_group,Dobrev:1977qv}, denoted here as
\begin{equation}\label{define_highest_weight_orthogonal}
    \vec{f}=(f_{1},f_{2},...,f_{p}),
\end{equation}
where
\begin{align}
 \label{highest_weight_so(2p)}   &f_{1} \geq f_{2} \geq ... \geq f_{p-1} \geq |f_{p}|, \hspace{12mm} \text{for spin}(2p),\\
  & f_{1} \geq f_{2} \geq ... \geq f_{p-1} \geq f_{p}\geq 0, \hspace{6mm} \text{for spin}(2p+1). \label{highest_weight_so(2p+1)}
\end{align}
The labels $f_{j}$ ($j=1,...,p$) in eqs.~(\ref{highest_weight_so(2p)}) and (\ref{highest_weight_so(2p+1)}) are all integers or all half-odd integers.
For spin$(2p)$, the label $f_{p}$ can be negative, while the representation $(f_{1},...,f_{p-1},-f_{p})$ is known as the `mirror image' of $(f_{1},...,,f_{p-1},f_{p})$ - see, e.g. Ref.~\cite{Todorov_1978}. For spin$(2p+1)$, any representation $\vec{f}$ is equivalent to its mirror image~\cite{Todorov_1978}.

The quadratic Casimir for the representation $\vec{f}=(f_{1},...,f_{p})$ is given by~\cite{Dobrev:1977qv}
\begin{align}
    c_{2}\left(\vec{f}\right)&=\sum_{j=1}^{p}f_{j}(f_{j}+2p-2j), \hspace{12mm} \text{for spin}(2p),\label{Casimir_spin(2p)}\\
  c_{2}\left(\vec{f}\right)&=\sum_{j=1}^{p}f_{j}(f_{j}+2p+1-2j), \hspace{6mm} \text{for spin}(2p+1). \label{Casimir_spin(2p+1)}
\end{align}


\noindent \textbf{UIR's of spin$\bm{(2p,1)}$ (even $\bm{D=2p \geq 4)}$.}  A UIR of spin$(2p,1)$ is specified by the set of labels $\vec{F}=(F_{0},F_{1},...,F_{p-1})$. The labels $F_{1},...,F_{p-1}$ satisfy
\begin{align}
    F_{1}\geq F_{2} \geq ... \geq F_{p-1} \geq 0
\end{align}
and they are all integers or all half-odd integers. A representation $(f_{1},...,f_{p})$ of spin$(2p)$ that is contained in the UIR $(F_{0},F_{1},...,F_{p-1})$ satisfies
\begin{align} \label{branching_rules_spin(2p,1)->spin(2p)}
  f_{1}\geq F_{1}\geq f_{2} \geq F_{2} \geq ...\geq f_{p-1} \geq F_{p-1} \geq |f_{p}| . 
\end{align}
Ottoson's labels~\cite{Ottoson} and our labels are related to each other by~\cite{Yale_Thesis}:
\begin{subequations}\label{Ottoson_notat_so(2p,p)}
\begin{align}
  f_{j}&=l_{2p-1,j}  +j-p,\hspace{5mm} (j=1,...,p),\\
   F_{j}&=l_{2p,j}  +j-p,\hspace{5mm} (j=1,...,p-1),\\
   F_{0}&=l_{2p,p}-p.
\end{align}
\end{subequations}
Schwarz's labels~\cite{Schwarz} and our labels are related to each other by:
\begin{subequations}\label{Schwarz_notat_so(2p,p)}
\begin{align}
  f_{j}&=m_{2p,p-j+1},\hspace{5mm} (j=1,...,p),\\
   F_{j}&=m_{2p+1,p-j}, \hspace{5mm} (j=1,...,p-1),\\
   F_{0}&=z_{2p+1,p}.
\end{align}
\end{subequations}

The UIR's of spin$(2p,1)$ (even $D=2p \geq 4$) are classified as follows:
\begin{itemize}
    \item \textbf{Principal Series} $\bm{D}_{\textbf{prin}}\bm{(\,\vec{F}\,)}$\textbf{:}
    \begin{align}\label{Principal_UIR_even}
        F_{0}=-p+\frac{1}{2}+i y=-\frac{D-1}{2}+iy, \hspace{5mm} (y>0).
    \end{align}
     The labels $F_{1},F_{2},...,F_{p-1}$ are all integers or all half-odd integers.

    \item \textbf{Complementary Series} $\bm{D}_{\textbf{comp}}\bm{(\,\vec{F}\,):}$ 
    \begin{align}\label{Compl_UIR_even}
     -\frac{D-1}{2}=-p+\frac{1}{2}\leq F_{0}<-\tilde{n}, \hspace{5mm}(\tilde{n}~ \text{is an integer and}~ 0 \leq \tilde{n} \leq p-1).   
    \end{align}
     If $0 \leq \tilde{n} < p-1$, then $F_{\tilde{n}+1}=F_{\tilde{n}+2}=...=F_{p-1}=0$ and $F_{1}, F_{2},...,F_{\tilde{n}}$ are all positive integers, while for the spin$(2p)$ content we have $f_{\tilde{n}+2}=f_{\tilde{n}+3}=...=f_{p}=0$. If $\tilde{n} = p-1$, then $F_{1},F_{2},...,F_{p-1}$ are all positive integers.~\footnote{Our Complementary Series is called Exceptional Series $D(e;l_{2p,1},...,l_{2p,p})$ in Ottoson's classification~\cite{Ottoson}. Also, our notation for the Complementary Series is related to Schwarz's notation~\cite{Schwarz} as follows. The case with $0 \leq \tilde{n} < p-1$ corresponds to $D^{k}(m_{2p+1,k+1}~...~m_{2p+1,p-1};x_{2p+1,p})$, where $k$ is related to $\tilde{n}$ by $k=p-\tilde{n}-1$, while the case with $\tilde{n}=p-1$ corresponds to $D^{0}(m_{2p+1,1}~...~m_{2p+1,p-1};x_{2p+1,p})$.}

     \item \textbf{Exceptional Series} $\bm{D}_{\textbf{ex}}\bm{(\,\vec{F}\,):}$ 
     \begin{align}\label{Exceptional_UIR_even}
        F_{0}=-\tilde{n}, \hspace{5mm}(\tilde{n}~ \text{is an integer and}~ 1\leq \tilde{n} \leq p-1).  
     \end{align}
     If $1 \leq \tilde{n} < p-1$, then $F_{\tilde{n}+1}=F_{\tilde{n}+2}=...=F_{p-1}=0$ and $F_{1}, F_{2},...,F_{\tilde{n}}$ are all positive integers, while for the spin$(2p)$ content we have $f_{\tilde{n}+1}=f_{\tilde{n}+2}=...=f_{p}=0$. If $\tilde{n} = p-1$, then $F_{1},F_{2},...,F_{p-1}$ are all positive integers, while $f_{p}=0$.~\footnote{Our Exceptional Series is called Supplementary Series $D(s;l_{2p,1},...,l_{2p,p})$ in Ottoson's classification~\cite{Ottoson,Yale_Thesis}. Also, our notation is related to Schwarz's notation~\cite{Schwarz} as follows. The case with $1 \leq \tilde{n} < p-1$ corresponds to $D^{k}(m_{2p+1,k+1}~...~m_{2p+1,p-1};m_{2p+1,p})$, where Schwarz's label $k$ is related to our label $\tilde{n}$ by $k=p-\tilde{n}-1$, while the case with $\tilde{n}=p-1$ corresponds to $D^{0}(m_{2p+1,1}~...~m_{2p+1,p-1};m_{2p+1,p})$.}

 \item \textbf{Discrete Series} $\bm{D}^{\bm{\pm}}\bm{(\,\vec{F}\,):}$ 
$F_{0}$ is real and it is an integer or half-odd integer at the same time as the labels $F_{1},F_{2},...,F_{p-1}$.\footnote{Our Discrete Series ${D}^{\pm}{(\vec{F})}$ are called Exceptional Series $D(\pm; l_{2p,1},...,l_{2p,p})$ in Ottoson's classification~\cite{Ottoson, Yale_Thesis}. Also, our Discrete Series ${D}^{\pm}{(\vec{F})}$ correspond to $D^{\pm}(m_{2p+1,1}\,...\,m_{2p+1,p-1};m_{2p+1,p})$ in Schwarz's classification~\cite{Schwarz}.} Also, the following conditions have to be satisfied:
\begin{align}\label{condition_for_discrete_series_+}
   F_{p-1} \geq f_{p}  \geq F_{0}+p \geq \frac{1}{2}\hspace{9mm}\text{for}~D^{+}(\,\vec{F}\,),
\end{align}
\begin{align}\label{condition_for_discrete_series_-}
  -F_{p-1} \leq f_{p} \leq -(F_{0}+p )\leq -\frac{1}{2}\hspace{6mm}\text{for}~D^{-}(\,\vec{F}\,).
  \end{align}
\end{itemize}
For a UIR of spin$(2p,1)$ labelled by $\vec{F}=(F_{0},F_{1},...,F_{p-1})$ the quadratic Casimir $C_{2}(\vec{F})$ is expressed as
\begin{align}\label{Casimir_Spin(2p,1)}
    C_{2}(\vec{F})=\sum_{k=0}^{p-1}F_{k}\,(F_{k}+2p-2k-1).
\end{align}
{This expression for the quadratic Casimir can be readily obtained by applying the ``analytic continuation'' techniques described in Refs.~\cite{Schwarz, Wong} to the quadratic Casimir~(\ref{Casimir_spin(2p+1)}) of spin$(2p+1)$. These techniques ``analytically continue'' $2p$ of the rotation generators of spin$(2p+1)$ to the $2p$ boost generators of spin$(2p,1)$ - for more details see Refs.~\cite{Schwarz, Wong}.}

\noindent \textbf{Note.} In the present paper, following Schwarz~\cite{Schwarz} and Ottoson~\cite{Ottoson}, for even $D$ the value $F_{0}=-(D-1)/2$ is not included in the Principal Series UIR's, but it is included in the Discrete Series UIR's instead. For odd $D$, the value $F_{0}=-(D-1)/2$ is included in the Principal Series UIR's in the present paper. However, in Ref.~\cite{Mixed_Symmetry_dS} the value $F_{0}=-(D-1)/2$ (corresponding to the weight $\Delta_{c}= (D-1)/2$) is included in the Principal Series UIR's for arbitrary $D$. The present note is important for reasons of clarity, as we are going to show that the spin-3/2 and spin-5/2 fields on even-dimensional $dS_{D}$ with mass parameter $M=0$ have $F_{0}= -(D-1)/2$ and they correspond to the Discrete Series UIR's in our paper (i.e. Principal Series in Ref.~\cite{Mixed_Symmetry_dS}) - see Section~\ref{Sec_Dictionary}. 

\noindent\textbf{UIR's of spin$\bm{(2p+1,1)}$ (odd $\bm{D=2p+1 \geq 3)}$.}  A UIR of spin$(2p+1,1)$ is labelled by $\vec{F}=(F_{0},F_{1},...,F_{p})$. The labels $F_{1},...,F_{p}$ satisfy
\begin{align}
    F_{1}\geq F_{2} \geq ... \geq F_{p} \geq 0
\end{align}
and they are all integers or half-odd integers. A representation $(f_{1},...,f_{p})$ of spin$(2p+1)$ that is contained in the UIR $\vec{F}=(F_{0},F_{1},...,F_{p})$ satisfies
\begin{align}\label{branching_rules_spin(2p+1,1)->spin(2p+1)}
  f_{1}\geq F_{1}\geq f_{2} \geq F_{2} \geq ...\geq f_{p} \geq F_{p} \geq 0 . 
\end{align}
Ottoson's labels~\cite{Ottoson} and our labels are related to each other by~\cite{Yale_Thesis}:
\begin{subequations}\label{Ottoson_notat_so(2p+1,p)}
\begin{align}
  f_{j}&=l_{2p,j}  +j-p-1\hspace{5mm} (j=1,...,p),\\
   F_{j}&=l_{2p+1,j}  +j-p\hspace{5mm} (j=1,...,p),\\
   F_{0}&=l_{2p+1,p+1}-p,
\end{align}
\end{subequations}
while Schwarz's labels~\cite{Schwarz} and our labels are related to each other by:
\begin{subequations}\label{Schwarz_notat_so(2p+1,p)}
\begin{align}
  f_{j}&=m_{2p+1,p-j+1}\hspace{5mm} (j=1,...,p),\\
   F_{j}&=m_{2p+2,p-j+1} \hspace{5mm} (j=1,...,p),\\
   F_{0}&=z_{2p+2,p+1}.
\end{align}
\end{subequations}

The UIR's of spin$(2p+1,1)$ (odd $D=2p+1 \geq 3$) are classified as follows:
\begin{itemize}
    \item \textbf{Principal Series} $\bm{D}_{\textbf{prin}}\bm{(\,\vec{F}\,):}$ 
    \begin{align}\label{Principal_UIR_odd}
       F_{0}=-p+i y=-\frac{D-1}{2}+iy, \hspace{5mm} (y\in \mathbb{R}). 
    \end{align}
     The labels $F_{1},F_{2},...,F_{p}$ are all integers or half-odd integers. If $F_{p}=0$, then the UIR with $F_{0}=-(D-1)/2+iy$ and the UIR with $F_{0}=-(D-1)/2-iy$ are equivalent, and thus we can let $y \geq 0$.
    
    \item \textbf{Complementary Series} $\bm{D}_{\textbf{comp}}\bm{(\,\vec{F}\,) :}$ 
    \begin{align}\label{Compl_UIR_odd}
        -\frac{D-1}{2}=-p< F_{0}<-\tilde{n}, \hspace{5mm}(\tilde{n}~ \text{is an integer and}~ 0 \leq \tilde{n} \leq p-1),
    \end{align}
    while $F_{\tilde{n}+1}=F_{\tilde{n}+2}=...=F_{p}=0$ and $F_{1}, F_{2},...,F_{\tilde{n}}$ are all positive integers, where for the spin$(2p+1)$ content we have $f_{\tilde{n}+2}=f_{\tilde{n}+3}=...=f_{p}=0$.~\footnote{Our Complementary Series corresponds to $D^{k}(m_{2p+2,k+1}~...~m_{2p+2,p};x_{2p+2,p+1})$ in Schwarz's classification~\cite{Schwarz}, where $k$ is related to $\tilde{n}$ by $k=p-\tilde{n}$.}

     \item \textbf{Exceptional Series} $\bm{D}_{\textbf{ex}}\bm{(\,\vec{F}\,):}$ 
     \begin{align}\label{Exceptional_UIR_odd}
          F_{0}=-\tilde{n}, \hspace{5mm}(\tilde{n}~ \text{is an integer and}~ 1\leq \tilde{n} \leq p-1),
     \end{align}
 where $F_{\tilde{n}+1}=F_{\tilde{n}+2}=...=F_{p}=0$ and $F_{1}, F_{2},...,F_{\tilde{n}}$ are all positive integers, where for the spin$(2p+1)$ content we have $f_{\tilde{n}+1}=f_{\tilde{n}+2}=...=f_{p}=0$.~\footnote{Our Exceptional Series corresponds to $D^{k}(m_{2p+2,k+1}~...~m_{2p+2,p};m_{2p+2,p+1})$ in Schwarz's classification~\cite{Schwarz}, where $k$ is related to $\tilde{n}$ by $k=p-\tilde{n}$.}
\end{itemize}
For a UIR of spin$(2p+1,1)$ specified by $\vec{F}=(F_{0},F_{1},...,F_{p})$ the quadratic Casimir $C_{2}(\vec{F})$ is expressed as\footnote{This expression for the quadratic Casimir can be obtained in the same way as in the even-dimensional case - see eq.~(\ref{Casimir_Spin(2p,1)}).}
\begin{align}\label{Casimir_Spin(2p+1,1)}
    C_{2}(\vec{F})=\sum_{k=0}^{p}F_{k}\,(F_{k}+2p-2k).
\end{align}

 \section{Spin-3/2 and spin-5/2 eigenmodes on \texorpdfstring{$dS_{D}$}{dS}}\label{Section_dS_eigenmodes}
 In this Section, we will obtain the spin-3/2 and spin-5/2 TT eigenmodes on global $dS_{D}$ satisfying eq.~(\ref{Dirac_eqn_fermion_dS}) using the method of separation of variables - see, e.g., Refs.~\cite{STSHS,Camporesi,CHH}. Schematically, in this method the spin-$(r+1/2)$ eigenmodes, $\Psi_{\mu_{1}...\mu_{r}}(t ,\bm{\theta_{D-1}})$, are expressed as products of two `parts'; namely a part describing the time-dependence (corresponding to a function of $t$) and another part describing the $\bm{\theta_{D-1}}$-dependence of the eigenmode (corresponding to tensor-spinor eigenmodes of the Dirac operator on $S^{D-1}$). In view of the classification of the spin$(D,1)$ UIR's under the decomposition spin$(D,1) \supset$ spin$(D)$, expressing our eigenmodes on $dS_{D}$ in terms of eigentensor-spinors on $S^{D-1}$ offers an easy way to understand the spin$(D)$ content of our dS eigenmodes. The outline of this Section is:
 \begin{itemize}
     \item In Subsection~\ref{Sub_Sec_eigenmodes_spheres}, we review the necessary material concerning the (totally symmetric) TT tensor-spinor eigenmodes of the Dirac operator on $S^{D-1}$ and the way they form representations of spin$(D)$~\cite{Homma}.
     \item In Subsections~\ref{Sub_Sec_separate_spin3/2_even} and \ref{Sub_Sec_separate_spin3/2_odd}, we present the construction of spin-3/2 TT eigenmodes on $dS_{D}$ in order to illustrate the method of separation of variables for tensor-spinor fields. Some basic results are tabulated in Tables 1 and 2.
     \item In Subsection~\ref{Sub_Sec_separate_spin5/2}, we summarise our main results concerning the spin-5/2 TT eigenmodes on $dS_{D}$.
 \end{itemize}
 
 \subsection{Tensor-spinor eigenmodes of the Dirac operator on \texorpdfstring{$S^{D-1}$}{SD-1} and representations of spin\texorpdfstring{$(D)$}{D}} \label{Sub_Sec_eigenmodes_spheres}
The spectrum of the Dirac operator acting on tensor-spinor eigenmodes on spheres, as well as the representations of spin$(D)$ formed by the eigenmodes, have been discussed in Refs.~\cite{Trautman_1993, Camporesi, Homma, CHH} (see also references therein).

Let $\tilde{\slashed{\nabla}} \equiv \tilde{\gamma}^{k} \tilde{\nabla}_{k}$ be the Dirac operator on $S^{D-1}$, where $\tilde{\nabla}_{j}$ is the covariant derivative on $S^{D-1}$. We are interested in rank-$\tilde{r} \geq 0$ totally symmetric TT tensor-spinor eigenmodes $\tilde{\psi}^{(\ell; \underline{m})}_{\pm \tilde{\mu}_{1} \tilde{\mu}_{2} ... \tilde{\mu}_{\tilde{r}}} (\bm{\theta_{D-1}})$ on $S^{D-1}$. The eigenmodes $\tilde{\psi}^{(\ell; \underline{m})}_{\pm \tilde{\mu}_{1} \tilde{\mu}_{2} ... \tilde{\mu}_{\tilde{r}}} (\bm{\theta_{D-1}})$ satisfy
\begin{align}
   & \tilde{\slashed{\nabla}}\tilde{\psi}^{(\ell; \underline{m})}_{\pm \tilde{\mu}_{1} \tilde{\mu}_{2} ... \tilde{\mu}_{\tilde{r}}}= \pm i \left(\ell+\frac{D-1}{2}\right)  \tilde{\psi}^{(\ell; \underline{m})}_{\pm \tilde{\mu}_{1} \tilde{\mu}_{2} ... \tilde{\mu}_{\tilde{r}}}\label{tensor-spinoreigen_SD-1} \\
   &\tilde{\gamma}^{\tilde{\mu}_{1}}\tilde{\psi}^{(\ell; \underline{m})}_{\pm \tilde{\mu}_{1} \tilde{\mu}_{2} ... \tilde{\mu}_{\tilde{r}}}=\tilde{\nabla}^{\tilde{\mu}_{1}}\tilde{\psi}^{(\ell; \underline{m})}_{\pm \tilde{\mu}_{1} \tilde{\mu}_{2} ... \tilde{\mu}_{\tilde{r}}}=0  \label{tensor-spinoreigen_SD-1_TT},
\end{align}
where the angular momentum quantum number on $S^{D-1}$, $\ell$, is allowed to take integer values with $\ell \geq \tilde{r}$. The two sets of eigenmodes, $\{ \tilde{\psi}^{(\ell; \underline{m})}_{+ \tilde{\mu}_{1} \tilde{\mu}_{2} ... \tilde{\mu}_{\tilde{r}}}  \}$ [with eigenvalue $+i (\ell +\tfrac{D-1}{2})$] and $\{ \tilde{\psi}^{(\ell; \underline{m})}_{- \tilde{\mu}_{1} \tilde{\mu}_{2} ... \tilde{\mu}_{\tilde{r}}}  \}$ [with eigenvalue $-i (\ell +\tfrac{D-1}{2})$], separately form representations of spin$(D)$. The label $\underline{m}$ represents quantum numbers (other than $\ell$) the values of which specify the content of the spin$(D)$ representation concerning the chain of subalgebras spin$(D-1) \, \supset $ spin$(D-2) \supset ...  \supset$ spin$(2)$.

\noindent \textbf{Odd $\bm{D \geq 3}$ (even-dimensional spheres).} For each allowed value of $\ell$ we have a representation of spin$(D)$ acting on the space of the eigenmodes $\{ \tilde{\psi}^{(\ell; \underline{m})}_{+ \tilde{\mu}_{1} \tilde{\mu}_{2} ... \tilde{\mu}_{\tilde{r}}}  \}$ (or $\{ \tilde{\psi}^{(\ell; \underline{m})}_{-\tilde{\mu}_{1} \tilde{\mu}_{2} ... \tilde{\mu}_{\tilde{r}}}  \}$) on $S^{D-1}$ with highest weight~(\ref{define_highest_weight_orthogonal}) given by~\cite{Homma}
 \begin{align}\label{highest_weight_odd_spin(D)_tensor-spinors}
     \vec{f}_{\tilde{r}}=\left(\ell +\frac{1}{2},\tilde{r}+\frac{1}{2},\frac{1}{2},...,\frac{1}{2} \right), \hspace{6mm}(\ell=\tilde{r},\tilde{r}+1,...),
 \end{align}
 where we have used the subscript $\tilde{r}$ in order to denote the `spin' of the representation - e.g.  $\vec{f}_{0}$ corresponds to a spinor representation, $\vec{f}_{1}$ to a TT vector-spinor representation, $\vec{f}_{2}$ to a rank-2 (totally symmetric) TT tensor-spinor representation and so forth.
The two sets of eigenmodes, $\{ \tilde{\psi}^{(\ell; \underline{m})}_{+ \tilde{\mu}_{1} \tilde{\mu}_{2} ... \tilde{\mu}_{\tilde{r}}}  \}$ and $\{ \tilde{\psi}^{(\ell; \underline{m})}_{- \tilde{\mu}_{1} \tilde{\mu}_{2} ... \tilde{\mu}_{\tilde{r}}}  \}$, form equivalent representations. For $D=5$ the highest weight is $\vec{f}_{\tilde{r}}= (\ell+1/2,  \tilde{r}+ 1/2)$. On $S^{2}$ - i.e. for $D=3$ - there are no totally symmetric TT eigenmodes satisfying eq.~(\ref{tensor-spinoreigen_SD-1}) with rank $\tilde{r} \geq 1$ - see Refs.~\cite{CHH, Letsios_in_progress, Letsios_arxiv_long} and Appendix~\ref{Appendix_tensor-spinors_twosphere}. However, eigenmodes with $\tilde{r}=0$ - i.e. eigenspinors $\tilde{\psi}^{(\ell; \underline{m})}_{\pm}$ of the Dirac operator~\cite{Camporesi} - exist on $S^{2}$ and the corresponding spin$(3)$ representation is labelled by the one-component highest weight $\ell+1/2$ (with $\ell = 0,1,...$).

\noindent \textbf{Even $\bm{D \geq 4}$ (odd-dimensional spheres).} For each allowed value of $\ell$ the eigenmodes $\{ \tilde{\psi}^{(\ell; \underline{m})}_{+ \tilde{\mu}_{1} \tilde{\mu}_{2} ... \tilde{\mu}_{\tilde{r}}}  \}$ on $S^{D-1}$ form a spin$(D)$ representation 
with highest weight~(\ref{define_highest_weight_orthogonal}) given by~\cite{Homma}
 \begin{align}\label{highest_weight_even+_spin(D)_tensor-spinors}
     \vec{f}^{+}_{\tilde{r}}=\left(\ell +\frac{1}{2},\tilde{r}+\frac{1}{2},\frac{1}{2},...,\frac{1}{2} \right), \hspace{6mm}(\ell=\tilde{r},\tilde{r}+1,...),
 \end{align}
 while the eigenmodes $\{ \tilde{\psi}^{(\ell; \underline{m})}_{- \tilde{\mu}_{1} \tilde{\mu}_{2} ... \tilde{\mu}_{\tilde{r}}}  \}$ form a representation
 with highest weight~\cite{Homma}
 \begin{align}\label{highest_weight_even-_spin(D)_tensor-spinors}
     \vec{f}^{-}_{\tilde{r}}=\left(\ell +\frac{1}{2},\tilde{r}+\frac{1}{2},\frac{1}{2},...,\frac{1}{2},-\frac{1}{2} \right), \hspace{6mm}(\ell=\tilde{r},\tilde{r}+1,...).
 \end{align}
 For $D=4$ the highest weights corresponding to the eigenmodes $\{ \tilde{\psi}^{(\ell; \underline{m})}_{\pm \tilde{\mu}_{1} \tilde{\mu}_{2} ... \tilde{\mu}_{\tilde{r}}}  \}$ are $\vec{f}^{\pm}_{\tilde{r}}= \left(\ell+1/2, \pm (\tilde{r}+1/2) \right)$.

For both even $D$ [eqs.~(\ref{highest_weight_even+_spin(D)_tensor-spinors}) and (\ref{highest_weight_even-_spin(D)_tensor-spinors})] and odd $D$ [eq.~(\ref{highest_weight_odd_spin(D)_tensor-spinors})], if the aforementioned irreducible representations of spin$(D)$ are contained in a spin$(D,1)$ representation, then the allowed values for the angular momentum quantum number $\ell$ might not just be $\ell = \tilde{r}, \tilde{r}+1,...$; $\ell$ might have to satisfy extra conditions because of the branching rules~(\ref{branching_rules_spin(2p,1)->spin(2p)}) and~(\ref{branching_rules_spin(2p+1,1)->spin(2p+1)}). This will become clear in the next Subsection as $\ell$ will have to satisfy $\ell \geq r$, where $r$ is the rank of the tensor-spinor eigenmodes on $dS_{D}$.



\subsection{Separating variables for spin-3/2 eigenmodes on \texorpdfstring{$dS_{D}$}{dS} for even \texorpdfstring{${D} \geq 4$}{D}} \label{Sub_Sec_separate_spin3/2_even}
Let us illustrate the method of separation of variables for the TT vector-spinor field $\Psi_{\mu} = (\Psi_{t}, \Psi_{\theta_{D-1}}, \Psi_{\theta_{D-2}},..., \Psi_{\theta_{1}})$ with arbitrary mass parameter $M$ on global $dS_{D}$ for even $D \geq 4$. 

The Dirac equation~(\ref{Dirac_eqn_fermion_dS}) is expressed as
\begin{align}
  & \left( \frac{\partial}{\partial{t}}+\frac{D+1}{2}\tanh{t} \right) \gamma^{t}\Psi_{t}+\frac{1}{\cosh{t}} \begin{pmatrix} 0 & i\tilde{\slashed{\nabla}} \\
    - i\tilde{\slashed{\nabla}} & 0
     \end{pmatrix}\Psi_{t} = -M \Psi_{t}, \label{Dirac_op_on_psi_t_dS} \\
    & \left( \frac{\partial}{\partial t}+\frac{D-3}{2}\tanh{t} \right) \gamma^{t}\Psi_{\theta_{j}}+\frac{1}{\cosh{t}} \begin{pmatrix} 0 & i\tilde{\slashed{\nabla}} \\
    - i\tilde{\slashed{\nabla}} & 0
     \end{pmatrix}\Psi_{\theta_{j}}- \tanh{t} \, \gamma_{\theta_{j}}\Psi_{t} =  -M \Psi_{\theta_{j}}, \label{Dirac_op_on_psi_theta_i_dS}
 \end{align}
 ($j=1,2,...,D-1$), where we have made use of eqs.~(\ref{covariant_deriv_vector_spinor}), (\ref{even_gammas}) and (\ref{Christoffels_dS})-(\ref{spin_connection_dS}), while $\gamma_{\theta_{j}}=e_{\theta_{j}}^{\hspace{2mm}k} \gamma_{k}$. There are two different ways in which we can separate variables for the TT vector-spinor $\Psi_{\mu}(t, \bm{\theta_{D-1}})$ giving rise to two different types of eigenmodes: the type-I modes and the type-II modes. These two different types of eigenmodes correspond to spin$(D)$ representations with different spin. In particular, the spin$(D)$ content that is relevant to type-I modes corresponds to the spinor representation $\vec{f}^{\pm}_{0} = (\ell + \tfrac{1}{2}, \tfrac{1}{2},...,\tfrac{1}{2}, \pm \tfrac{1}{2})$ with $\ell=1,2,...\,$.\footnote{Under the decomposition spin$(D,1)$ $\supset$ spin$(D)$, the branching rules~(\ref{branching_rules_spin(2p,1)->spin(2p)}) give rise to the restriction $\ell \geq 1$. One can also arrive at this restriction on $\ell$ by requiring the regularity of type-I eigenmodes, as we will discuss below. See Refs.~\cite{Letsios_in_progress, Letsios_arxiv_long} for more details concerning the explicit form of the eigenmodes.} The spin$(D)$ content that is relevant to type-II modes corresponds to the vector-spinor representation $\vec{f}^{\pm}_{1} = (\ell + \tfrac{1}{2}, \tfrac{3}{2}, \tfrac{1}{2},...,\tfrac{1}{2}, \pm \tfrac{1}{2})$ with $\ell=1,2,...$.  

\noindent \textbf{Type-I modes.} Let us denote the type-I modes with spin$(D)$ content given by $\vec{f}^{\pm}_{0} = (\ell + \tfrac{1}{2}, \tfrac{1}{2},...,\tfrac{1}{2}, \pm \tfrac{1}{2})$ as $\Psi_{\mu}^{(M;\,\tilde{r}=0 ,\,\pm \ell; \,\underline{m})} (t, \bm{\theta_{D-1}})$, where the label $\underline{m}$ has the same meaning as in Subsection~\ref{Sub_Sec_eigenmodes_spheres}. We start with the case of $\vec{f}^{-}_{0} = (\ell + \tfrac{1}{2}, \tfrac{1}{2},...,\tfrac{1}{2}, - \tfrac{1}{2})$, i.e. with the type-I modes $\Psi_{\mu}^{(M;\,\tilde{r}=0 ,\,- \ell; \,\underline{m})} (t, \bm{\theta_{D-1}})$. As in Refs.~\cite{Camporesi, CHH, Letsios}, we separate variables for the $t$-component by expressing it in terms of upper and lower spinor components, as
\begin{equation}\label{TYPE_I_thetaN_negative_spin_3/2_anal_cont}
    {\Psi}^{(M;\,\tilde{r}=0 ,\,- \ell; \,\underline{m})}_{t}(t,\bm{\theta}_{D-1})= \begin{pmatrix} -i \varPhi^{(1)}_{M \ell}(t) \, \tilde{\psi}_{-}^{(\ell; \underline{m})} (\bm{\theta_{D-1}})  \\ -  \varPsi^{(1)}_{M \ell}(t) \, \tilde{\psi}_{-}^{(\ell; \underline{m})} (\bm{\theta_{D-1}})  \end{pmatrix},
\end{equation}
where $\tilde{\psi}_{-}^{(\ell; \underline{m})}$ are the $2^{D/2-1}$-dimensional eigenspinors of $\tilde{\slashed{\nabla}}$ on $S^{D-1}$ [see Eq.~(\ref{tensor-spinoreigen_SD-1})]. Now, we have to determine the functions of time $\varPhi^{(1)}_{M \ell}(t)$ and $\varPsi^{(1)}_{M \ell}(t)$ - the superscript `(1)' in these functions has been used for later convenience. By substituting eq.~(\ref{TYPE_I_thetaN_negative_spin_3/2_anal_cont}) into the Dirac equation~(\ref{Dirac_op_on_psi_t_dS}), we can eliminate the lower component in eq.~(\ref{TYPE_I_thetaN_negative_spin_3/2_anal_cont}). We find in this manner the second order equation for $\varPhi^{(1)}_{M \ell}(t)$ 
\begin{align}\label{diff_equation_for_phi_a=1}
  \mathcal{D}_{(1)} \varPhi^{(1)}_{M \ell} =M^{2} \,\varPhi^{(1)}_{M \ell},
  \end{align}
where the differential operator $\mathcal{D}_{(1)}$ is a special case of the following family of differential operators: 
   \begin{align}\label{diff_operator}
      \mathcal{D}_{(a)} =&~\frac{\partial^{2}}{\partial x^{2}}+(D+2a-1)\cot{x}\frac{\partial}{\partial x}+\left(\ell+\frac{D-1}{2}\right)\frac{\cos{x}}{\sin^{2}{x}} \nonumber\\
      &-\frac{(\ell+\frac{D-1}{2})^{2}-\frac{1}{4}{(D+2a-1)(D+2a-3)}}{\sin^{2}{x}} -\frac{(D+2a-1)^{2}}{4},
  \end{align}
where we have defined
\begin{align} \label{useful_variable_x(t)}
    x= x(t) := \frac{\pi}{2} - i t
\end{align}
with $\cos{x}= i \sinh{t}$ and $\sin{x}=\cosh{t}$. For later convenience, instead of just solving the eigenvalue equation~(\ref{diff_equation_for_phi_a=1}), we can solve the more general equation
\begin{align}\label{diff_equation_for_phi_a}
  \mathcal{D}_{(a)} \varPhi^{(a)}_{M \ell} =M^{2} \,\varPhi^{(a)}_{M \ell},
  \end{align}
for arbitrary integer $a$. The solution is given by
\begin{align}
    {\varPhi}^{(a)}_{M \ell}(t) = & \left(\cos{\frac{x(t)}{2}}\right)^{\ell+1-a}\left(\sin{\frac{x(t)}{2}}\right)^{\ell-a}\nonumber\\
    & \times F\left(-i{M}+\frac{D}{2}+\ell,i{M}+\ell+\frac{D}{2};\ell+\frac{D}{2};\sin^{2}\frac{x(t)}{2}\right)\label{phi_aM_t},
\end{align}
where $F(A, B; C; z)$ is the Gauss hypergeometric function~\cite{gradshteyn2007}, while
\begin{align}
   &\cos{\frac{x(t)}{2}}=\left(  \sin{\frac{x(t)}{2}} \right)^{*}=\frac{\sqrt{2}}{2}\,\left(\cosh{\frac{t}{2}} + i \sinh{\frac{t}{2}} \right)    \label{cosx/2}.
       \end{align}
Thus, we have now determined the upper component of ${\Psi}^{(M;\,\tilde{r}=0 ,\,- \ell; \,\underline{m})}_{t}$ in eq.~(\ref{TYPE_I_thetaN_negative_spin_3/2_anal_cont}), where $\varPhi^{(1)}_{M \ell}$ is given by eq.~(\ref{phi_aM_t}) with $a=1$.

In order to determine the lower component in eq.~(\ref{TYPE_I_thetaN_negative_spin_3/2_anal_cont}), we substitute eq.~(\ref{TYPE_I_thetaN_negative_spin_3/2_anal_cont}) into the Dirac equation~(\ref{Dirac_op_on_psi_t_dS}) and we straightforwardly find the relations
 \begin{align}
    \left(\frac{d}{dt}+\frac{D+1}{2}\tanh{t}-\frac{i\left( \ell+\frac{D-1}{2} \right)}{\cosh{t}}\,\right){\varPsi}^{(1)}_{{M} \ell}(t)&=- {M}\,{\varPhi}^{(1)}_{M \ell}(t) \label{psia=1_to_phi_sphere_analcont},\end{align}
\begin{align}\label{phia=1_to_psi_sphere_analcont}
  \left(\frac{d}{d t}+\frac{D+1}{2}\tanh{t}+\frac{i \left(\ell+\frac{D-1}{2} \right)}{\cosh{t}} \,\right)\varPhi^{(1)}_{M \ell}(t)&= \, M \,\varPsi^{(1)}_{M \ell}(t) .
  \end{align}
 Then, substituting eq.~(\ref{phi_aM_t}) (with $a=1$) into eq.~(\ref{phia=1_to_psi_sphere_analcont}) and using well-known properties of the hypergeometric function~\cite{gradshteyn2007}, we find
\begin{align}
   {\varPsi}^{(1)}_{M \ell}(t)=&~\frac{-i {M}}{\ell+\frac{D}{2}}  \left(\cos{\frac{x(t)}{2}}\right)^{\ell-1} \left(\sin{\frac{x(t)}{2}} \right)^{\ell} \nonumber\\ &\times F\left(-iM+\frac{D}{2}+\ell, iM+\ell+\frac{D}{2};\ell+\frac{D+2}{2};\sin^{2}\frac{x(t)}{2}\right).\label{psi_a=1M_t}
\end{align}
 For later convenience, let us note that ${\varPsi}^{(1)}_{M \ell}(t)$ corresponds to a special case (i.e. the case with $a=1$) of the following functions:
\begin{align}
   \varPsi^{(a)}_{M \ell}(t)
    =&~\frac{- i {M}}{\ell+\frac{D}{2}}  \left(\cos{\frac{x(t)}{2}}\right)^{\ell-a} \left(\sin{\frac{x(t)}{2}} \right)^{\ell+1-a} \nonumber\\ &\times F\left(-i{M}+\frac{D}{2}+\ell, i{M}+\ell+\frac{D}{2};\ell+\frac{D+2}{2};\sin^{2}\frac{x(t)}{2}\right).\label{psi_aM_t}
\end{align}
These functions solve the differential equation~$ (\hat{\mathcal{D}}_{(a)}-M^{2}) \varPsi^{(a)}_{M \ell}(t)=0$ where the differential operator $\hat{\mathcal{D}}_{(a)}$ is given by eq.~(\ref{diff_operator}) with $x$ replaced by $\pi - x$.
Thus, we have now also determined the lower component of ${\Psi}^{(M;\,\tilde{r}=0 ,\,- \ell; \,\underline{m})}_{t}$ in eq.~(\ref{TYPE_I_thetaN_negative_spin_3/2_anal_cont}).

Now, by following the same procedure as the one described above, we can separate variables for the type-I modes $  {\Psi}^{(M;\,\tilde{r}=0 ,\,+ \ell; \,\underline{m})}_{\mu}(t,\bm{\theta_{D-1}})$ corresponding to the spin$(D)$ highest weight $\vec{f}^{+}_{0} = (\ell + \tfrac{1}{2}, \tfrac{1}{2},...,\tfrac{1}{2})$. We find 
\begin{equation}\label{TYPE_I_thetaN_positive_spin_3/2_anal_cont}
    {\Psi}^{(M;\,\tilde{r}=0 ,\,+ \ell; \,\underline{m})}_{t}(t,\bm{\theta_{D-1}})= \begin{pmatrix}  \varPsi^{(1)}_{M \ell}(t) \, \tilde{\psi}_{+}^{(\ell; \underline{m})} (\bm{\theta_{D-1}})  \\ i  \varPhi^{(1)}_{M \ell}(t) \, \tilde{\psi}_{+}^{(\ell; \underline{m})} (\bm{\theta_{D-1}})  \end{pmatrix}.
\end{equation}
 The rest of the vector components of the type-I modes, ${\Psi}^{(M;\,\tilde{r}=0 ,\, \pm \ell; \,\underline{m})}_{\theta_{j}}$ ($j=1,...,D-1$), can be straightforwardly determined by substituting the known expressions for ${\Psi}^{(M;\,\tilde{r}=0 ,\,\pm \ell; \,\underline{m})}_{t}$ [eqs.~(\ref{TYPE_I_thetaN_negative_spin_3/2_anal_cont}) and (\ref{TYPE_I_thetaN_positive_spin_3/2_anal_cont})] into the TT conditions~(\ref{TT_conditions_fermions_dS}). By doing so, one finds that there is a proportionality factor of $\tfrac{1}{\ell}$ in the expressions for each of the ${\Psi}^{(M;\,\tilde{r}=0 ,\,\pm \ell; \,\underline{m})}_{\theta_{j}}$ and, thus, the regularity of type-I eigenmodes gives rise to the restriction $\ell \geq 1$. However, here we will not present explicit expressions for ${\Psi}^{(M;\,\tilde{r}=0 ,\,\pm \ell; \,\underline{m})}_{\theta_{j}}$ ($j=1,...,D-1$) as they are lengthy and they are not needed for our analysis. The interested reader can find the explicit expressions in Refs.~\cite{Letsios_in_progress, Letsios_arxiv_long}.

\noindent \textbf{Type-II modes.} Let us denote the type-II modes with spin$(D)$ content given by $\vec{f}^{\pm}_{1} = (\ell + \tfrac{1}{2}, \tfrac{3}{2},\tfrac{1}{2},...,\tfrac{1}{2}, \pm \tfrac{1}{2})$ as $\Psi_{\mu}^{(M;\,\tilde{r}=1 ,\,\pm \ell; \,\underline{m})} (t, \bm{\theta_{D-1}})$ ($\ell \geq 1$). The type-II modes are TT vector-spinors on $S^{D-1}$ and thus ${\Psi}^{(M;\,\tilde{r}=1 ,\,\pm \ell; \,\underline{m})}_{t}(t,\bm{\theta}_{D-1}) =0$. The components ${\Psi}^{(M;\,\tilde{r}=1 ,\,\pm \ell; \,\underline{m})}_{\theta_{j}}(t,\bm{\theta}_{D-1})$ can be determined by applying the method of separation of variables as in the case of the type-I modes. However, now we have to express ${\Psi}^{(M;\,\tilde{r}=1 ,\,\pm \ell; \,\underline{m})}_{\theta_{j}}(t,\bm{\theta}_{D-1})$ in terms of TT eigenvector-spinors on $S^{D-1}$, instead of eigenspinors on $S^{D-1}$. By applying the method of separation of variables to the Dirac equation~(\ref{Dirac_op_on_psi_theta_i_dS}), we find 
\begin{equation}\label{TYPE_II_negative_spin_3/2_anal_cont}
  {\Psi}^{(M;\,\tilde{r}=1 ,\,- \ell; \,\underline{m})}_{t} (t, \bm{\theta_{D-1}})=0, \hspace{5mm} {\Psi}^{(M;\,\tilde{r}=1 ,\,- \ell; \,\underline{m})}_{\theta_{j}}(t,\bm{\theta}_{D-1})= \begin{pmatrix}  \varPhi^{(-1)}_{M \ell}(t) \, \tilde{\psi}_{- \theta_{j}}^{(\ell; \underline{m})} (\bm{\theta_{D-1}})  \\ - i \varPsi^{(-1)}_{M \ell}(t) \, \tilde{\psi}_{- \theta{j}}^{(\ell; \underline{m})} (\bm{\theta_{D-1}})  \end{pmatrix}
\end{equation}
and
\begin{equation}\label{TYPE_II_positive_spin_3/2_anal_cont}
  {\Psi}^{(M;\,\tilde{r}=1 ,\,+ \ell; \,\underline{m})}_{t}(t,\bm{\theta}_{D-1}) =0, \hspace{5mm} {\Psi}^{(M;\,\tilde{r}=1 ,\,+ \ell; \,\underline{m})}_{\theta_{j}}(t,\bm{\theta}_{D-1})= \begin{pmatrix} i  \varPsi^{(-1)}_{M \ell}(t) \, \tilde{\psi}_{+ \theta_{j}}^{(\ell; \underline{m})} (\bm{\theta_{D-1}})  \\ - \varPhi^{(-1)}_{M \ell}(t) \, \tilde{\psi}_{+ \theta{j}}^{(\ell; \underline{m})} (\bm{\theta_{D-1}})  \end{pmatrix},
\end{equation}
($j = 1, ..., D-1$) where $\tilde{\psi}_{\pm \theta{j}}^{(\ell; \underline{m})} (\bm{\theta_{D-1}})$ are the TT eigenvector-spinors~(\ref{tensor-spinoreigen_SD-1}) on $S^{D-1}$. The functions $\varPhi^{(-1)}_{M \ell}(t)$ and $\varPsi^{(-1)}_{M \ell}(t)$ are given by eqs.~(\ref{phi_aM_t}) and (\ref{psi_aM_t}), respectively, with $a=-1$.
 
 \noindent \textbf{Summary.} Some basic results concerning the spin-3/2 eigenmodes for even $D \geq 4$ are tabulated in Table~1.
 
\subsection{Separating variables for spin-3/2 eigenmodes on \texorpdfstring{${dS_{D}}$}{dS} for odd \texorpdfstring{${{D} \geq 3}$}{D}} \label{Sub_Sec_separate_spin3/2_odd}
The Dirac equation~(\ref{Dirac_eqn_fermion_dS}) is expressed as
\begin{align}
  & \left( \frac{\partial}{\partial{t}}+\frac{D+1}{2}\tanh{t} \right) \gamma^{t}\Psi_{t}+\frac{1}{\cosh{t}} \tilde{\slashed{\nabla}} \Psi_{t} = -M \Psi_{t}, \label{Dirac_op_on_psi_t_dS_odd} \\
    & \left( \frac{\partial}{\partial t}+\frac{D-3}{2}\tanh{t} \right) \gamma^{t}\Psi_{\theta_{j}}+\frac{1}{\cosh{t}} \tilde{\slashed{\nabla}}\Psi_{\theta_{j}}- \tanh{t} \, \gamma_{\theta_{j}}\Psi_{t} =  -M \Psi_{\theta_{j}}, \label{Dirac_op_on_psi_theta_i_dS_odd}
 \end{align}
 ($j=1,2,...,D-1$), where the gamma matrices are now given by eq.~(\ref{odd_gammas}). As in the even-dimensional case, we have two different types of eigenmodes depending on their spin$(D)$ content.
 
 \noindent \textbf{Type-I modes.} Let us denote the type-I modes with spin$(D)$ content given by $\vec{f}_{0} = (\ell + \tfrac{1}{2}, \tfrac{1}{2},...,\tfrac{1}{2})$ as $\Psi_{\mu}^{(M;\,\tilde{r}=0 , \ell; \,\underline{m})} (t, \bm{\theta_{D-1}})$ (with $\ell \geq 1$). As in Refs.~\cite{Camporesi, CHH, Letsios}, we separate variables as
 \begin{align}\label{TYPE_I_thetaN_Nodd_spin_3/2}
  \Psi_{t}^{(M;\,\tilde{r}=0 , \ell; \,\underline{m})} (t, \bm{\theta_{D-1}})  &=\frac{1}{\sqrt{2}}(\bm{1}+ \gamma^{t}) \left\{-i\,\varPhi^{(1)}_{M \ell}(t) +i  \, \varPsi^{(1)}_{M \ell}(t)\gamma^{t}\right\}  \tilde{\psi}_{-}^{(\ell; \underline{m})} (\bm{\theta_{D-1}}),
\end{align}
where $\tilde{\psi}_{-}^{(\ell; \underline{m})}$ are the eigenspinors~(\ref{tensor-spinoreigen_SD-1}) on $S^{D-1}$, while $i \tilde{\psi}_{+}^{(\ell; \underline{m})} = \gamma^{t}\tilde{\psi}_{-}^{(\ell; \underline{m})}$ as $\gamma^{t}$ anti-commutes with $\tilde{\slashed{\nabla}}$. Substituting eq.~(\ref{TYPE_I_thetaN_Nodd_spin_3/2}) into the Dirac equation~(\ref{Dirac_op_on_psi_t_dS_odd}), we find that $\varPhi^{(1)}_{M \ell}(t) , \varPsi^{(1)}_{M \ell}(t)$ must satisfy the relations~(\ref{psia=1_to_phi_sphere_analcont}) and (\ref{phia=1_to_psi_sphere_analcont}). Then, we readily find that $\varPhi^{(1)}_{M \ell}(t)$ is given by eq.~(\ref{phi_aM_t}) with $a=1$, while  $\varPsi^{(1)}_{M \ell}(t)$ is given by eq.~(\ref{psi_a=1M_t}). The components $\Psi_{\theta_{j}}^{(M;\,\tilde{r}=0 , \ell; \,\underline{m})} (t, \bm{\theta_{D-1}})$ can be determined with the use of the TT conditions~(\ref{TT_conditions_fermions_dS}), as in the even-dimensional case.

\noindent \textbf{Type-II modes.} The type-II modes $\Psi_{\mu}^{(M;\,\tilde{r}=1 , \ell; \,\underline{m})} (t, \bm{\theta_{D-1}})$ correspond to the following spin$(D)$ representation: $\vec{f}_{1} = (\ell + \tfrac{1}{2}, \tfrac{3}{2},\tfrac{1}{2},...,\tfrac{1}{2})$ with $\ell \geq 1$ and they exist for $D>3$. We separate variables as
 \begin{align}\label{TYPE_II_Nodd_spin_3/2}
 \Psi_{t}^{(M;\,\tilde{r}=1 , \ell; \,\underline{m})} (t, \bm{\theta_{D-1}})  &= 0  , \nonumber \\
  \Psi_{\theta_{j}}^{(M;\,\tilde{r}=1 , \ell; \,\underline{m})} (t, \bm{\theta_{D-1}})  &=\frac{1}{\sqrt{2}}(\bm{1}+ \gamma^{t}) \left\{\,\varPhi^{(-1)}_{M \ell}(t)   - \varPsi^{(-1)}_{M \ell}(t)\gamma^{t}\right\}  \tilde{\psi}_{-\theta_{j}}^{(\ell; \underline{m})} (\bm{\theta_{D-1}}),
\end{align}
$(j=1,...,D-1)$ where $\tilde{\psi}_{- \theta_{j}}^{(\ell; \underline{m})}$ are the eigenvector-spinors~(\ref{tensor-spinoreigen_SD-1}) on $S^{D-1}$, while $i \tilde{\psi}_{+ \theta_{j}}^{(\ell; \underline{m})} = \gamma^{t}\tilde{\psi}_{- \theta_{j}}^{(\ell; \underline{m})}$. Substituting eq.~(\ref{TYPE_II_Nodd_spin_3/2}) into the Dirac equation~(\ref{Dirac_op_on_psi_theta_i_dS_odd}), we find that $\varPhi^{(-1)}_{M \ell}(t)$ and $\varPsi^{(-1)}_{M \ell}(t)$ are given by eqs.~(\ref{phi_aM_t}) and (\ref{psi_aM_t}), respectively, with $a=-1$.

\noindent \textbf{Summary.} Some basic results concerning the spin-3/2 eigenmodes for odd $D \geq 3$ are tabulated in Table 2.
\subsection{Spin-5/2 eigenmodes on \texorpdfstring{$dS_{D}$}{dS}}\label{Sub_Sec_separate_spin5/2}
In the case of rank-2 totally symmetric tensor-spinors $\Psi_{\mu \nu}$ - which satisfy eqs.~(\ref{Dirac_eqn_fermion_dS}) and (\ref{TT_conditions_fermions_dS}) with $r=2$ on $dS_{D}$ - the method of separation of variables can be applied in a way analogous to the case of TT vector-spinors. Depending on the spin$(D)$ content of the spin-5/2 dS eigenmode we can distinguish three types of modes: type-I, type-II and type-III modes (the last two exist for $D>3$). Here we will just summarise some basic results for the TT spin-5/2 eigenmodes on $dS_{D}$. Below we use the same notation for the labels of the eigenmodes as in the spin-3/2 case, while we refer again to the spin$(D)$ content of the eigenmodes using the highest weights $\vec{f}_{\tilde{r}}^{\pm}$ for even $D$ [eqs.~(\ref{highest_weight_even+_spin(D)_tensor-spinors}) and (\ref{highest_weight_even-_spin(D)_tensor-spinors})] and $\vec{f}_{\tilde{r}}$ for odd $D$ [eq.~(\ref{highest_weight_odd_spin(D)_tensor-spinors})].
 
 \noindent  \textbf{Even $\bm{D \geq 4}$.} The TT spin-5/2 eigenmodes on $dS_{D}$ and their spin$(D)$ content are given by:
\begin{align}\label{TYPE_I_thetaNthetaN_positive_and_negative_spin_5/2}
  &\textbf{Type-I:}~ \vec{f}^{\pm}_{0}= \left(\ell+ \frac{1}{2}, \frac{1}{2},...,\frac{1}{2},\pm \frac{1}{2} \right), \hspace{4mm} \ell=2,3,...\, \nonumber\\ &{\Psi}^{(M;\,\tilde{r}=0 ,\,-\ell; \,\underline{m})}_{tt}= \begin{pmatrix}  -\varPhi^{(2)}_{M \ell}(t) \, \tilde{\psi}_{-}^{(\ell; \underline{m})} (\bm{\theta_{D-1}})  \\ i \varPsi^{(2)}_{M \ell}(t) \, \tilde{\psi}_{-}^{(\ell; \underline{m})} (\bm{\theta_{D-1}})  \end{pmatrix},
    {\Psi}^{(M;\,\tilde{r}=0 ,\,+ \ell; \,\underline{m})}_{tt}= \begin{pmatrix} -i \varPsi^{(2)}_{M \ell}(t) \, \tilde{\psi}_{+}^{(\ell; \underline{m})} (\bm{\theta_{D-1}})  \\  \varPhi^{(2)}_{M \ell}(t) \, \tilde{\psi}_{+}^{(\ell; \underline{m})} (\bm{\theta_{D-1}})  \end{pmatrix}.
\end{align}
\begin{align}\label{TYPE_II_thetaNthetaj_positive_and_negative_spin_5/2}
 &\textbf{Type-II:} ~\vec{f}^{\pm}_{1}= \left(\ell+ \frac{1}{2}, \frac{3}{2},\frac{1}{2},...,\frac{1}{2},\pm \frac{1}{2} \right), \hspace{4mm} \ell=2,3,...\, \nonumber\\
 &{\Psi}^{(M;\,\tilde{r}=1 ,\,- \ell; \,\underline{m})}_{t t}= {\Psi}^{(M;\,\tilde{r}=1 ,\,+ \ell; \,\underline{m})}_{t t}=0 ,\nonumber \\   &{\Psi}^{(M;\,\tilde{r}=1 ,\,- \ell; \,\underline{m})}_{t \theta_{j}}= \begin{pmatrix}  -i\varPhi^{(0)}_{M \ell}(t) \, \tilde{\psi}_{- \theta_{j}}^{(\ell; \underline{m})} (\bm{\theta_{D-1}})  \\ -  \varPsi^{(0)}_{M \ell}(t) \, \tilde{\psi}_{- \theta_{j}}^{(\ell; \underline{m})} (\bm{\theta_{D-1}})  \end{pmatrix},~{\Psi}^{(M;\,\tilde{r}=1 ,\,+ \ell; \,\underline{m})}_{t \theta_{j}}= \begin{pmatrix}  \varPsi^{(0)}_{M \ell}(t) \, \tilde{\psi}_{+ \theta_{j}}^{(\ell; \underline{m})} (\bm{\theta_{D-1}})  \\ i  \varPhi^{(0)}_{M \ell}(t) \, \tilde{\psi}_{+ \theta_{j}}^{(\ell; \underline{m})} (\bm{\theta_{D-1}})  \end{pmatrix}.
\end{align}
\begin{align}\label{TYPE_III_thetajthetak_positive_and_negative_spin_5/2}
 &\textbf{Type-III:} ~ \vec{f}^{\pm}_{2}= \left(\ell+ \frac{1}{2}, \frac{5}{2},\frac{1}{2},...,\frac{1}{2},\pm \frac{1}{2} \right), \hspace{4mm} \ell=2,3,... \, \nonumber\\
 &{\Psi}^{(M;\,\tilde{r}=2 ,\,- \ell; \,\underline{m})}_{t \mu}= {\Psi}^{(M;\,\tilde{r}=2 ,\,+ \ell; \,\underline{m})}_{t \mu}=0,~ \hspace{5mm} \nonumber \\  
 &{\Psi}^{(M;\,\tilde{r}=2 ,\,- \ell; \,\underline{m})}_{ \theta_{j} \theta_{k}}= \begin{pmatrix}  \varPhi^{(-2)}_{M \ell}(t) \, \tilde{\psi}_{- \theta_{j} \theta_{k}}^{(\ell; \underline{m})} (\bm{\theta_{D-1}})  \\ - i \varPsi^{(-2)}_{M \ell}(t) \, \tilde{\psi}_{- \theta_{j} \theta_{k}}^{(\ell; \underline{m})} (\bm{\theta_{D-1}})  \end{pmatrix},{\Psi}^{(M;\,\tilde{r}=2 ,\,+ \ell; \,\underline{m})}_{ \theta_{j} \theta_{k}}= \begin{pmatrix}  i\varPsi^{(-2)}_{M \ell}(t) \, \tilde{\psi}_{+ \theta_{j} \theta_{k}}^{(\ell; \underline{m})} (\bm{\theta_{D-1}})  \\ -  \varPhi^{(-2)}_{M \ell}(t) \, \tilde{\psi}_{+ \theta_{j} \theta_{k}}^{(\ell; \underline{m})} (\bm{\theta_{D-1}})  \end{pmatrix},
\end{align}
where $\tilde{\psi}_{\pm \theta_{j} \theta_{k}}^{(\ell; \underline{m})}$ are the rank-2 tensor-spinor eigenmodes~(\ref{tensor-spinoreigen_SD-1}) on $S^{D-1}$, while $\mu = t, \theta_{D-1},...,\theta_{2}, \theta_{1}$ and $j,k=1,...,D-1$. The components that have not been written down explicitly can be found from the TT conditions~(\ref{TT_conditions_fermions_dS}) (for explicit expressions for all the components see Refs.~\cite{Letsios_in_progress, Letsios_arxiv_long}).

 \noindent  \textbf{Odd $\bm{D \geq 3}$.} The TT spin-5/2 eigenmodes on $dS_{D}$ and their spin$(D)$ content are given by:
\begin{align}\label{TYPE_I_thetaNthetaN_odd_spin_5/2}
  &\textbf{Type-I:}~ \vec{f}_{0}= \left(\ell+ \frac{1}{2}, \frac{1}{2},...,\frac{1}{2} \right), \hspace{4mm} \ell=2,3,...\,\nonumber\\
  &{\Psi}^{(M;\,\tilde{r}=0 ,\,\ell; \,\underline{m})}_{tt}=\frac{1}{\sqrt{2}}(\bm{1}+ \gamma^{t}) \left\{\,-\varPhi^{(2)}_{M \ell}(t)   + \varPsi^{(2)}_{M \ell}(t)\gamma^{t}\right\}  \tilde{\psi}_{-}^{(\ell; \underline{m})} (\bm{\theta_{D-1}}).
\end{align}
\begin{align}\label{TYPE_II_thetaNthetaj_odd_spin_5/2}
 &\textbf{Type-II (for $D>3$)}: ~\vec{f}_{1}= \left(\ell+ \frac{1}{2}, \frac{3}{2},\frac{1}{2},...,\frac{1}{2} \right), \hspace{4mm} \ell=2,3,...\, \nonumber\\
 &{\Psi}^{(M;\,\tilde{r}=1 , \ell; \,\underline{m})}_{t t}=0, \nonumber \\  
 &{\Psi}^{(M;\,\tilde{r}=1 ,\,\ell; \,\underline{m})}_{t\theta_{j}}=\frac{1}{\sqrt{2}}(\bm{1}+ \gamma^{t}) \left\{\,-i\varPhi^{(0)}_{M \ell}(t)   + i\varPsi^{(0)}_{M \ell}(t)\gamma^{t}\right\}  \tilde{\psi}_{-\theta_{j}}^{(\ell; \underline{m})} (\bm{\theta_{D-1}}).
\end{align}
\begin{align}\label{TYPE_III_thetajthetak_odd_spin_5/2}
 &\textbf{Type-III (for $D>3$):} ~ \vec{f}_{2}= \left(\ell+ \frac{1}{2}, \frac{5}{2},\frac{1}{2},...,\frac{1}{2}\right), \hspace{4mm} \ell=2,3,...  \nonumber\\
 &{\Psi}^{(M;\,\tilde{r}=2 ,\, \ell; \,\underline{m})}_{t \mu}=0, \nonumber \\  
 &{\Psi}^{(M;\,\tilde{r}=2 ,\,\ell; \,\underline{m})}_{\theta_{j}\theta_{k}}=\frac{1}{\sqrt{2}}(\bm{1}+ \gamma^{t}) \left\{\,\varPhi^{(-2)}_{M \ell}(t)   - \varPsi^{(-2)}_{M \ell}(t)\gamma^{t}\right\}  \tilde{\psi}_{-\theta_{j} \theta_{k}}^{(\ell; \underline{m})} (\bm{\theta_{D-1}}),
\end{align}
where $\mu = t, \theta_{D-1},...,\theta_{2}, \theta_{1}$ and $j,k=1,...,D-1$. As in the even-dimensional case, the components that have not been written down explicitly can be found from the TT conditions~(\ref{TT_conditions_fermions_dS}).
\section{Quadratic Casimir for spin-3/2 and spin-5/2 eigenmodes on \texorpdfstring{$dS_{D}$}{dSD}} \label{Sec_quadratic_Casimir}
In order to find the values of the spin$(D,1)$ quadratic Casimir corresponding to the representation formed by our spin-3/2 and spin-5/2 eigenmodes we will use the ``analytic continuation'' techniques that have been already used in Refs.~\cite{STSHS, Letsios}. More specifically, we will use the fact that $dS_{D}$ can be obtained by an “analytic continuation” of $S^{D}$. The line element of $S^{D}$ can be written as
\begin{align}
    d\Omega_{D}^{2}= d \theta^{2}_{D} + \sin^{2}{\theta_{D}} \, d \Omega^{2}_{D-1} ,
\end{align}
where $0 \leq \theta_{D} \leq \pi $. By replacing the angle $\theta_{D}$ in $d\Omega_{D}^{2}$ as:
 \begin{align}\label{coord_change_analytic_cont}
     \theta_{D} \rightarrow x(t)= \frac{\pi}{2} - i t,
 \end{align}
 ($t \in \mathbb{R}$) we find the line element~(\ref{dS_metric}) for global $dS_{D}$ ($x(t)$ coincides with the `useful' variable that we have already introduced in eq.~(\ref{useful_variable_x(t)})).

\noindent \textbf{Quadratic Casimir for tensor-spinor eigenmodes on $\bm{S^{D}}$.} Motivated by the aforementioned observation, we can obtain the field equations (\ref{Dirac_eqn_fermion_dS}) and (\ref{TT_conditions_fermions_dS}) for spin-$(r+1/2)$ fields on $dS_{D}$ by analytically continuing the equations for totally symmetric TT tensor-spinors of rank $r$ on $S^{D}$:
 \begin{align}
   &\slashed{\nabla}\psi_{\pm\mu_{1}...\mu_{r}}=\pm i \left(n+\frac{D}{2} \right) \psi_{\pm\mu_{1}...\mu_{r}} ,\hspace{5mm}(n=r,\,r+1,...) \label{Dirac_eqn_fermion_SN_minus}\\
   & \nabla^{\alpha}\psi_{\pm\alpha \mu_{2}...\mu_{r}}=0, \hspace{4mm}  \gamma^{\alpha}\psi_{\pm\alpha \mu_{2}...\mu_{r}}=0 \label{TT_conditions_fermions_SN_minus},
\end{align}
where $\psi_{\pm \mu_{1}...\mu_{r}}$ is a tensor-spinor on $S^{D}$, while $n$ is the angular momentum quantum number on $S^{D}$.
Equations~(\ref{Dirac_eqn_fermion_SN_minus}) and~(\ref{TT_conditions_fermions_SN_minus}) are essentially the $D$-dimensional counterparts of eqs.~(\ref{tensor-spinoreigen_SD-1}) and (\ref{tensor-spinoreigen_SD-1_TT}), while now $n$ on $S^{D}$ plays the role of $\ell$ on $S^{D-1}$. As we discussed in Subsection~\ref{Sub_Sec_eigenmodes_spheres}, the spin$(D+1)$ representations formed by tensor-spinor eigenmodes of the Dirac operator on $S^{D}$ are known~\cite{Homma}. Using eqs.~(\ref{Casimir_spin(2p)}) and~(\ref{Casimir_spin(2p+1)}), the spin$(D+1)$ quadratic Casimir corresponding to the eigenmodes $\psi_{\pm \mu_{1} ... \mu_{r}}$ on $S^{D}$ is readily found to be
\begin{align}\label{quadratic_Casimir_Spin(N+1)_STSSH}
   \mathcal{C}^{(S^{D})}_{eigen}=& ~\left(n +\frac{D}{2} \right)^{2}-r-\frac{D(D-1)}{4}+\frac{(D-2)(D-3)}{8}+s(s+D-2) \\
   =&\,-\nabla^{\mu}\nabla_{\mu}+\frac{(D-2)(D-3)}{8}+s(s+D-2) \hspace{5mm} (\text{where}~ s=r+{1}/{2}), \nonumber
\end{align}
for all $D \geq 3$, while in the second line we used that $\nabla^{\mu}\nabla_{\mu}$ acts on $\psi_{\pm \mu_{1} ... \mu_{r}}$ as $$\nabla^{\mu}\nabla_{\mu}=\slashed{\nabla}^{2}+\frac{D(D-1)}{4}+r.$$

\noindent \textbf{Analytic continuation to $\bm{dS_{D}}$.}
Without loss of generality, we can choose to analytically continue the eigentensor-spinors with either one of the two signs for the eigenvalue in eq.~(\ref{Dirac_eqn_fermion_SN_minus}), since each of the two sets of modes, $\{\psi_{+\mu_{1}...\mu_{r}}\}$ and $\{\psi_{-\mu_{1}...\mu_{r}}\}$, forms independently a unitary representation of spin$(D+1)$ labelled by $n$ (see Subsection~\ref{Sub_Sec_eigenmodes_spheres}). Here we choose to analytically continue the eigentensor-spinors $\psi_{-\mu_{1}...\mu_{r}}$. We perform analytic continuation by making the following replacements in eqs.~(\ref{Dirac_eqn_fermion_SN_minus}) and (\ref{TT_conditions_fermions_SN_minus})\footnote{By making the replacements~(\ref{replacements}), the tensor-spinor $\psi_{-\mu_{1}...\mu_{r}}$ on $S^{D}$ is analytically continued to the tensor-spinor $\Psi_{\mu_{1}...\mu_{r}}$ [eq.~(\ref{Dirac_eqn_fermion_dS})] on $dS_{D}$. Alternatively, we could analytically continue the eigentensor-spinors on $S^{D}$ by making the replacement $\theta_{D} \rightarrow \pi/2 + it$ instead of the replacement~(\ref{coord_change_analytic_cont}). The analytically continued eigentensor-spinors with $\theta_{D} \rightarrow \pi/2 - it$ and the ones with $\theta_{D} \rightarrow \pi/2 + it$ are related to each other by charge conjugation. However, these two cases of eigenmodes form equivalent representations of spin$(D,1)$ - see Refs.~\cite{Letsios_in_progress, Letsios_arxiv_long}.}:
\begin{equation}\label{replacements}
        \theta_{D} \rightarrow x(t)=\frac{\pi}{2} -it , \hspace{10mm} n \rightarrow -i M  - \frac{D}{2}\hspace{10mm}(t\in \mathbb{R})
    \end{equation}
and we obtain eqs.~(\ref{Dirac_eqn_fermion_dS}) and (\ref{TT_conditions_fermions_dS}), respectively, for tensor-spinors $\Psi_{\mu_{1}...\mu_{r}}$ with mass parameter $M$ on $dS_{D}$. Recall that the values of interest for $M$ are: $M \in \mathbb{R}$ (corresponding to massive fermions of spin $s \geq 3/2$), as well as the purely imaginary values of $M$ corresponding to the strictly/partially massless tunings~(\ref{values_mass_parameter_masslessness_fermion}). The prescription for obtaining the explicit form of dS eigenmodes by analytically continuing eigenmodes on $S^{D}$ can be found in Refs.~\cite{STSHS, Letsios, Letsios_in_progress, Letsios_arxiv_long}.

\noindent \textbf{Quadratic Casimir for tensor-spinor eigenmodes on $\bm{dS_{D}}$.} With the use of the replacements~(\ref{replacements}), we analytically continue the quadratic Casimir on $S^{D}$~[Eq.~(\ref{quadratic_Casimir_Spin(N+1)_STSSH})], and we find the value of the quadratic Casimir on $dS_{D}$:
\begin{align}\label{quadratic_Casimir_Spin(N,1)_analcont_STSSH}
   \mathcal{C}^{(dS_{D})}_{eigen}=&~ -{M}^{2}-r-\frac{D(D-1)}{4}+\frac{(D-2)(D-3)}{8}+s(s+D-2) 
\end{align}
(with $s=r+1/2$), which holds for all $D \geq 3$ and for all totally symmetric TT tensor-spinor eigenmodes with spin $s \geq 1/2$ and mass parameter $M$ on $dS_{D}$. Specialising to the spin-3/2 TT eigenmodes we find
\begin{align}\label{quadratic_Casimir_Spin(N,1)_analcont_STSSH_spin3/2}
   \mathcal{C}^{(dS_{D})}_{eigen}=&~ -{M}^{2}-\frac{(D-1)(D-8)}{8},
\end{align}
while for the spin-5/2 TT eigenmodes we find
\begin{align}\label{quadratic_Casimir_Spin(N,1)_analcont_STSSH_spin5/2}
   \mathcal{C}^{(dS_{D})}_{eigen}=&~ -{M}^{2}-\frac{D(D-17)}{8}.
\end{align}
 \begin{table}
 \begin{center}
\begin{tabular}{ |c|c|c| } 
 \hline
 \textbf{Type of eigenmode} &  \textbf{Notation} & \textbf{spin$(\bm{D})$ content} \\ 
 Type-I & $\Psi_{\mu}^{(M;\,\tilde{r}=0 , \pm \ell; \,\underline{m})}$ & $\vec{f}^{\pm}_{0}= (\ell+ \frac{1}{2}, \frac{1}{2},...,\frac{1}{2},\pm \frac{1}{2} ), \hspace{2mm} \ell=1,2,...$ \\ 
 Type-II & $\Psi_{\mu}^{(M;\,\tilde{r}=1 , \pm \ell; \,\underline{m})}$ & $\vec{f}^{\pm}_{1}= (\ell+ \frac{1}{2}, \frac{3}{2},\frac{1}{2},...,\frac{1}{2},\pm \frac{1}{2} ), \hspace{2mm} \ell=1,2,...$ \\ 
 \hline
\end{tabular}
\caption{\textbf{Spin-3/2 TT eigenmodes with mass parameter $\bm{M}$ on $\bm{dS_{D}}$ 
(even $\bm{D \geq 4}$).} For real $M \neq 0$, type-I and type-II modes together form a spin$(D,1)$ Principal Series UIR. For $M=0$ the representation is reducible as the two sets of eigenmodes $\{\Psi_{\mu}^{(M;\,\tilde{r} , - \ell; \,\underline{m})}\}_{\tilde{r}=0,1}$ and $\{\Psi_{\mu}^{(M;\,\tilde{r} , + \ell; \,\underline{m})}\}_{\tilde{r}=0,1}$ separately form Discrete Series UIR's of spin$(D,1)$. For $M = \pm i (D-2)/2$ (strictly massless tuning) the type-I modes become pure gauge modes, while the type-II modes are the physical modes forming a non-unitary representation for $D \neq 4$ and a direct sum of two `chiral' UIR's in the Discrete Series for $D=4$ - see Section~\ref{Sec_Dictionary}. All these results have been also explained by studying the group-theoretic properties of the eigenmodes in Refs.~\cite{Letsios_in_progress, Letsios_arxiv_long}.}
\end{center}\label{table_spin-3/2_real_M_even}
\end{table}
 \begin{table}\label{table_spin-3/2_real_M_odd}
 \begin{center}
\begin{tabular}{ |c|c|c| } 
 \hline
 \textbf{Type of eigenmode} &  \textbf{Notation} & \textbf{spin$(\bm{D})$ content} \\ 
 Type-I & $\Psi_{\mu}^{(M;\,\tilde{r}=0 , \ell; \,\underline{m})}$ & $\vec{f}_{0}= (\ell+ \frac{1}{2}, \frac{1}{2},..., \frac{1}{2} ), \hspace{2mm} \ell=1,2,...$ \\ 
 Type-II (for $D >3$) & $\Psi_{\mu}^{(M;\,\tilde{r}=1 ,  \ell; \,\underline{m})}$ & $\vec{f}_{1}= (\ell+ \frac{1}{2}, \frac{3}{2},\frac{1}{2},..., \frac{1}{2} ), \hspace{2mm} \ell=1,2,...$ \\ 
 \hline
\end{tabular}
\caption{\textbf{Spin-3/2 TT eigenmodes with mass parameter $\bm{M}$ on $\bm{dS_{D}}$ 
(odd $\bm{D \geq 3}$)}. Type-II modes exist for $D>3$. For real $M$, type-I and type-II modes together form a Principal Series UIR of spin$(D,1)$ for $D>3$. For real $M$ and $D=3$, type-I modes form a Principal Series UIR of spin$(3,1)$. For $M = \pm i (D-2)/2$ (strictly massless tuning) and $D>3$, the type-I modes become pure gauge modes, while the type-II modes are the physical modes forming a non-unitary strictly massless representation. At the strictly massless tuning $M = \pm i /2$ for $D=3$, the type-I modes are again pure gauge modes and they form a non-unitary representation of spin$(3,1)$ - see Section~\ref{Sec_Dictionary}. All these results have been also explained by studying the group-theoretic properties of the eigenmodes in Refs.~\cite{Letsios_in_progress, Letsios_arxiv_long}.}
\end{center}
\end{table}


 \section{Strictly and partially massless representations: non-unitarity for \texorpdfstring{$D \neq 4$}{Dneq4} and unitarity for \texorpdfstring{$D  = 4$}{D=4}} \label{Sec_main_result_unitarity}
Here we will obtain the main result of this paper: the strictly massless spin-3/2 field, as well as the strictly and partially massless spin-5/2 fields, on $dS_{D}$ ($D \geq 3$) cannot be unitary  unless $D=4$. Note that we already know 
the values of the quadratic Casimir~[eqs.~(\ref{quadratic_Casimir_Spin(N,1)_analcont_STSSH_spin3/2}) and (\ref{quadratic_Casimir_Spin(N,1)_analcont_STSSH_spin5/2})] for the representations formed by our dS eigenmodes for any mass parameter $M$. By specialising to the strictly/partially massless tunings~(\ref{values_mass_parameter_masslessness_fermion}), we find
\begin{align}
   \mathcal{C}^{(dS_{D})}_{eigen}=&~ \frac{D(D+1)}{8},\hspace{8mm} \text{spin-3/2, strictly massless, } \label{casimir_spin3/2_str_massless} \\
   \mathcal{C}^{(dS_{D})}_{eigen}=&~ \frac{D(D+17)}{8},\hspace{8mm} \text{spin-5/2, strictly massless, }\label{casimir_spin5/2_str_massless} \\
   \mathcal{C}^{(dS_{D})}_{eigen}=&~ \frac{(D+1)(D+8)}{8},\hspace{5mm} \text{spin-5/2, partially massless. }\label{casimir_spin5/2_part_massless}
\end{align}
 Apart from the values of the quadratic Casimir~(\ref{casimir_spin3/2_str_massless})-(\ref{casimir_spin5/2_part_massless}), we also know the spin$(D)$ content of the spin$(D,1)$ representations formed by our dS eigenmodes - see Tables~1 and 2, as well as Subsections~\ref{Sub_Sec_separate_spin3/2_even}-\ref{Sub_Sec_separate_spin5/2}. Keeping these results in mind, we can use the classification of the UIR's in Section~\ref{Sec_Classification_UIRs} in order to readily deduce the (non-)unitarity of the representations formed by our strictly/partially massless eigenmodes on $dS_{D}$. 
 
 First, let us identify which types of dS eigenmodes correspond to pure gauge modes and which to physical modes in the strictly/partially massless theories. By `physical modes' we mean the eigenmodes that form the strictly/partially massless representation of spin$(D,1)$ and that correspond to the (non-gauge) propagating degrees of freedom of the theory. (If the representation formed by the eigenmodes is non-unitary, then the name `physical modes' could be misleading as the theory is, of course, unphysical due to the appearance of negative probabilities.) The pure gauge modes describe pure gauge degrees of freedom of the theory. If a dS invariant scalar product exists, then the pure gauge modes have zero norm and they are orthogonal to all physical modes~\cite{STSHS, Letsios_in_progress, Letsios_arxiv_long}. The generators of spin$(D,1)$ act in terms of the Lie-Lorentz derivative~(\ref{Lie_Lorentz}) on equivalence classes of physical modes with equivalence relation given by: ``For any two physical modes $\Psi^{(1)}_{\mu_{1}...\mu_{r}}$ and $\Psi^{(2)}_{\mu_{1}...\mu_{r}}$ we have $\Psi^{(1)}_{\mu_{1}...\mu_{r}} \sim \Psi^{(2)}_{\mu_{1}...\mu_{r}}$ if and only if their difference $\Psi^{(1)}_{\mu_{1}...\mu_{r}} -\Psi^{(2)}_{\mu_{1}...\mu_{r}}$ is a linear combination of pure gauge modes''.
 \subsection{Pure gauge modes and physical modes}\label{Subsection_puregauge_phys}
 \noindent \textbf{Pure gauge and physical modes for strictly massless spin-3/2 field.} The mass parameter for the strictly massless spin-3/2 field is given by $M=\pm i (D-2)/2 $ [this is found by letting $r=\uptau=1$ in eq.~(\ref{values_mass_parameter_masslessness_fermion})]. The spin-3/2 type-I modes are the pure gauge modes of the theory, while the spin-3/2 type-II modes are the physical modes that form the (strictly massless) representation of spin$(D,1)$. More specifically, we find that for $M=\pm i (D-2)/2 $ all type-I modes [see eqs.~(\ref{TYPE_I_thetaN_negative_spin_3/2_anal_cont}) and  (\ref{TYPE_I_thetaN_positive_spin_3/2_anal_cont}) for even $D \geq 4$ and eq.~(\ref{TYPE_I_thetaN_Nodd_spin_3/2}) for odd $D \geq 3$] are expressed in a pure gauge form as:
\begin{align}\label{PG_TYPEI_spin3/2}
    \Psi^{(PG)}_{\pm \mu}(t, \bm{\theta_{D-1}})=\left( \nabla_{\mu}\pm \frac{i}{2}\gamma_{\mu}  \right) \Lambda_{\pm}(t, \bm{\theta_{D-1}}),
\end{align}
where for convenience we have omitted all quantum number labels from $\Psi^{(PG)}_{\pm \mu}$ and $\Lambda_{\pm}$. The subscript `$\pm$' in $\Psi^{(PG)}_{\pm \mu}$ denotes the sign of the mass parameter $M =\pm i (D-2)/2$.  The spinor gauge functions $\Lambda_{\pm}(t, \bm{\theta_{D-1}})$ satisfy
\begin{align}
    \slashed{\nabla}\Lambda_{\pm}=\mp i \,\frac{D}{2}\,\Lambda_{\pm}.
\end{align}

   \noindent \textbf{Pure gauge and physical modes for strictly massless spin-5/2 field.} The mass parameter for the strictly massless spin-5/2 field is given by $M=\pm i D/2 $ [this is found by letting $r=2$ and $\uptau=1$ in eq.~(\ref{values_mass_parameter_masslessness_fermion})]. There are two types of pure gauge modes, namely the type-I and type-II modes. The spin-5/2 type-III modes are the physical modes that form the (strictly massless) representation of spin$(D,1)$. More specifically, we find that for $M=\pm i D/2 $ all type-I modes [see eq.~(\ref{TYPE_I_thetaNthetaN_positive_and_negative_spin_5/2}) for even $D \geq 4$ and eq.~(\ref{TYPE_I_thetaNthetaN_odd_spin_5/2}) for odd $D\geq 3$] and all type-II modes [see eq.~(\ref{TYPE_II_thetaNthetaj_positive_and_negative_spin_5/2}) for even $D \geq 4$ and eq.~(\ref{TYPE_II_thetaNthetaj_odd_spin_5/2}) for odd $D\geq 5$] are expressed in a pure gauge form as:
   \begin{align}\label{PG_TYPEB_spin5/2}
    {\Psi}^{(PG)}_{\pm \mu \nu}(t, \bm{\theta_{D-1}}) = \left(\nabla_{(\mu}\pm \frac{i}{2} \gamma_{(\mu} \right) \lambda_{\pm \nu)}(t, \bm{\theta}_{D-1})
\end{align}
for some TT vector-spinor gauge functions $\lambda_{\pm \mu}(t, \bm{\theta_{D-1}})$ with
\begin{align}
   & \slashed{\nabla}\lambda_{\pm \mu}= \mp i \,\frac{D+2}{2} \,\lambda_{\pm \mu}\\
   & \gamma^{\mu} \lambda_{\pm \mu}= \nabla^{\mu}  \lambda_{\pm \mu}=0.
\end{align}
(The gauge functions for type-I modes are different from the gauge functions for type-II modes - for more details see Refs.~\cite{Letsios_in_progress, Letsios_arxiv_long}.)

 \noindent \textbf{Pure gauge and physical modes for partially massless spin-5/2 field.} The mass parameter for the partially massless spin-5/2 field is given by $M=\pm i (D-2)/2 $ [this is found by letting $r=2$ and $\uptau=2$ in eq.~(\ref{values_mass_parameter_masslessness_fermion})]. The type-I modes are the pure gauge modes of the theory. Both type-II and type-III modes are physical modes that form the (partially massless) representation of spin$(D,1)$. For $M=\pm  i (D-2)/2$ all type-I modes are expressed in a pure gauge form as:
   \begin{align}\label{PG_TYPEI_spin5/2_partially}
   {\Psi}^{(PG)}_{ \pm \mu \nu}(t, \bm{\theta_{D-1}})= \left( \nabla_{(\mu}\nabla_{\nu)} \pm i  \gamma_{(\mu}\nabla_{\nu)} +\frac{3}{4} g_{\mu \nu} \right) \varphi_{\pm}(t, \bm{\theta_{D-1}}),
\end{align} 
where the spinor gauge functions $\varphi_{\pm}(t, \bm{\theta_{D-1}})$ satisfy
    \begin{align}
\slashed{\nabla}\varphi_{\pm}=\mp i\, \frac{D+2}{2}\,\varphi_{\pm}.
\end{align}
Explicit expressions on global $dS_{D}$ for the eigenmodes corresponding to the gauge functions in eqs.~(\ref{PG_TYPEI_spin3/2}), (\ref{PG_TYPEB_spin5/2}) and (\ref{PG_TYPEI_spin5/2_partially}) can be found in Refs.~\cite{Letsios_in_progress, Letsios_arxiv_long}.

\noindent \textbf{Remark 6.1.} On $dS_{3}$, both spin-$3/2$ and spin-$5/2$ theories with arbitrary mass parameters have only type-I modes. Thus, specialising to the strictly/partially massless theories on $dS_{3}$, we conclude that all eigenmodes for these theories are pure gauge modes.

\noindent \textbf{Remark 6.2.} In the fermionic strictly/partially massless theories of spin $s=r+1/2$ and depth $\uptau=1,...,r$ on global even-dimensional $dS_{D}$ ($ D \geq 4$), we can deduce which eigenmodes are pure gauge modes and which are physical modes from their spin$(D)$ content. The latter corresponds to the highest weights $\vec{f}^{+}_{\tilde{r}}= (\ell+1/2, \tilde{r}+1/2,1/2,...,1/2)$ and $\vec{f}_{\tilde{r}}^{-}= (\ell+1/2, \tilde{r}+1/2,1/2,...,1/2,-1/2)$ with $\tilde{r} \leq r \leq \ell$. The pure gauge modes correspond to the cases with $0 \leq \tilde{r} \leq r - \uptau$, while the physical modes correspond to $r - \uptau+1 \leq \tilde{r} \leq r$.

\noindent \textbf{Remark 6.3.} In the fermionic strictly/partially massless theories of spin $s=r+1/2$ and depth $\uptau=1,...,r$ on global odd-dimensional $dS_{D}$ ($ D \geq 3$), we can deduce which eigenmodes are pure gauge modes and which are physical modes from their spin$(D)$ content. The latter corresponds to the highest weights $\vec{f}_{\tilde{r}}= (\ell+1/2, \tilde{r}+1/2,1/2,...,1/2)$ with $\tilde{r} \leq r \leq \ell$. As in the even-dimensional case, the pure gauge modes correspond to the cases with $0 \leq \tilde{r} \leq r - \uptau$, while the physical modes correspond to $r - \uptau+1 \leq \tilde{r} \leq r$.

The validity of Remarks 6.1-6.3 for the spin-3/2 and spin-5/2 fields has been demonstrated in this paper, as well as in Refs.~\cite{Letsios_in_progress, Letsios_arxiv_long}. However, we expect that these remarks also hold for all strictly/partially massless fields with half-odd-integer spins $s \geq 3/2$. This expectation is also motivated by the well-studied case of totally symmetric tensors~\cite{STSHS}.

\subsection{Studying the (non-)unitarity of the strictly/partially massless theories with spin \texorpdfstring{$s =3/2, 5/2$}{s}} \label{Subsec_nonunitarity}

Our `tools' in order to demonstrate that the unitarity of the strictly/partially massless fields of spin $s=3/2, 5/2$ occurs only for $D=4$ are: on the one hand the values of the quadratic Casimir~[eqs.~(\ref{casimir_spin3/2_str_massless})-(\ref{casimir_spin5/2_part_massless})] and the spin$(D)$ content of the physical modes [see Tables 1 and 2 and Remarks 6.1-6.3] and, on the other hand, the classification of the UIR's in Section~\ref{Sec_Classification_UIRs}. 
Although the readers can readily convince themselves about the non-unitarity for $D \neq 4$ (given our aforementioned tools), we will present here a detailed discussion concerning the strictly massless spin-3/2 field. The cases of the strictly and partially massless spin-5/2 fields can then be treated in the same manner and, therefore, we will not present their details here.

\noindent \textbf{Non-unitarity for odd $\bm{D = 2p+1 \geq 5}$.} Let $\vec{F}= (F_{0},F_{1},...,F_{p})$ be the spin$(2p+1,1)$ representation formed by the physical spin-3/2 modes. The corresponding spin$(D)$ content is given by $\vec{f}_{1}= (\ell+1/2,3/2,1/2,...,1/2)$ with $\ell \geq 1$ - see Remark 6.3. The labels $F_{1},F_{2},...,F_{p}$ must all be half-odd-integers. It is clear that these values for $F_{1},...,F_{p}$ - as well as the spin$(D)$ content - correspond neither to the UIR's of the Exceptional Series~(\ref{Exceptional_UIR_odd}), nor to the UIR's of the Complementary Series~(\ref{Compl_UIR_odd}), since these UIR's allow only integer values for $F_{1},...,F_{p}$. Then, the only remaining candidate that could accommodate the strictly massless spin-3/2 field is the Principal Series~(\ref{Principal_UIR_odd}), where $F_{0}=-(D-1)/2+iy$ ($y \in \mathbb{R}$). We will readily show that the Principal Series \textit{cannot} accommodate the strictly massless spin-3/2 field. Suppose, for the sake of contradiction, that the strictly massless spin-3/2 representation $\vec{F}=(F_{0},..,F_{p})$ belongs to the Principal Series UIR's~(\ref{Principal_UIR_odd}). Since we already know the spin$(D)$ content of $\vec{F}$, by using the branching rules~(\ref{branching_rules_spin(2p+1,1)->spin(2p+1)}) we find that the following must hold: $F_{1}=3/2$, $F_{2} \in  \{1/2, 3/2 \}$ and $F_{3}=...=F_{p-1}=F_{p}=1/2$. Moreover, the quadratic Casimir for the Principal Series $C_{2}(\vec{F})$ [eq.~(\ref{Casimir_Spin(2p+1,1)})] must coincide with the quadratic Casimir $\mathcal{C}^{(dS_{D})}_{eigen}= \tfrac{D(D+1)}{8}$ [eq.~(\ref{casimir_spin3/2_str_massless})] corresponding to the physical modes. By equating these two values for the quadratic Casimir we find that $F_{0}$ must satisfy
\begin{align}\label{contradiction_intermediate}
    F_{0}(F_{0}+D-1)+F_{2}(F_{2}+D-5)+\frac{3}{2}=0, \hspace{5mm}\text{with} ~F_{2}\in \left \{\frac{1}{2},\frac{3}{2} \right \}.
\end{align}
For $F_{2}=1/2$ this equation gives $F_{0}=-1/2$ or $F_{0}= -D +3/2$, i.e. we arrive at a contradiction as these values for $F_{0}$ do \textit{not} correspond to the Principal Series  for odd $D \geq 5$. Similarly, for $F_{2}= 3/2$ we arrive again at a contradiction because eq.~(\ref{contradiction_intermediate}) gives $F_{0}=-3/2$ or $F_{0}= -D +5/2$ and these values do \textit{not} correspond to the Principal Series for odd $D \geq 5$. 
To conclude, we have proved that the strictly massless spin-3/2 field \textit{cannot} be accommodated by any UIR of spin$(D,1)$ for odd $D\geq 5$.

\noindent \textbf{Non-unitarity for $\bm{D = 3}$.} As we discussed earlier, on $dS_{3}$ the strictly massless spin-3/2 field (as well as the strictly and partially massless spin-5/2 fields) has only pure gauge modes - see Remark 6.1. However, it is worth showing here that the spin$(3,1)$ representation formed by the pure gauge modes of the strictly massless spin-3/2 theory is non-unitary. Let $\vec{F}= (F_{0},F_{1})$ be the spin$(3,1)$ representation formed by the pure gauge modes. The spin$(3)$ content for this representation is given by $\ell+1/2$ with $\ell \geq 1$ - see Remark 6.3. Also, the label $F_{1}$ must be a half-odd-integer. Thus, we can rule out both the Complementary Series~(\ref{Compl_UIR_odd}) and the Exceptional Series~(\ref{Exceptional_UIR_odd}) [in fact, the Exceptional Series does not exist for $D=3$~\cite{Schwarz}]. Now, as in the case with odd $D \geq 5$, it is easy to show that the quadratic Casimir for the spin$(3,1)$ Principal Series $C_{2}(\vec{F})$ [eq.~(\ref{Casimir_Spin(2p+1,1)})] does \textit{not} coincide with the field-theoretic quadratic Casimir $\mathcal{C}^{(dS_{3})}_{eigen}= \tfrac{3}{2}$ [eq.~(\ref{casimir_spin3/2_str_massless})] on $dS_{3}$.\footnote{For arbitrary $D$, the physical modes have the same value for the quadratic Casimir as the pure gauge modes.}

\noindent \textbf{Non-unitarity for even $\bm{D = 2p \geq 6}$.} Let $\vec{F}= (F_{0},F_{1},...,F_{p-1})$ be the spin$(2p,1)$ representation formed by the physical spin-3/2 modes. The corresponding spin$(D)$ content is given by $\vec{f}^{+}_{1}= (\ell+1/2,3/2,1/2,...,1/2)$ and $\vec{f}^{-}_{1}= (\ell+1/2,3/2,1/2,...,1/2,-1/2)$ with $\ell \geq 1$ (see Remark 6.1), while the labels $F_{1},F_{2},...,F_{p-1}$ must all be half-odd-integers. These values are incompatible with both the UIR's of the Exceptional Series~(\ref{Exceptional_UIR_even}) and the UIR's of the Complementary Series~(\ref{Compl_UIR_even}). Then, the UIR's that are still candidates for accommodating the strictly massless spin-3/2 field are: the Principal Series~(\ref{Principal_UIR_even}) and the Discrete Series~(\ref{condition_for_discrete_series_+}) and (\ref{condition_for_discrete_series_-}). Now, the following steps are as in the case with odd $D \geq 5$, i.e. we can prove by contradiction that the strictly massless spin-3/2 field corresponds neither to the Principal Series nor to the Discrete Series for even $D \geq 6$. In particular, starting with the contradicting assumption that $\vec{F}$ belongs to the Principal or Discrete Series, and making use of the branching rules~(\ref{branching_rules_spin(2p,1)->spin(2p)}), we equate the field-theoretic Casimir~(\ref{casimir_spin3/2_str_massless}) with the quadratic Casimir from the UIR's [eq.~(\ref{Casimir_Spin(2p,1)})]. By doing so, we find again that $F_{0}$ must satisfy eq.~(\ref{contradiction_intermediate}). Then, we readily arrive at a contradiction because the values of $F_{0}$ that satisfy eq.~(\ref{contradiction_intermediate}) agree neither with the Principal Series nor with the Discrete Series UIR's for even $D \geq 6$.
To conclude, we have proved that the strictly massless spin-3/2 field \textit{cannot} be accommodated by any UIR of spin$(D,1)$ for even $D\geq 6$.

\noindent \textbf{Unitarity for $\bm{D = 4}$.} The mass parameter for the strictly massless spin-3/2 field on $dS_{4}$ is $M= \pm i$. However, the physical modes with $M=i$ and the ones with $M=-i$ form equivalent representations\footnote{This can be readily understood as follows. If we act with $\gamma^{5}$ on any spin-3/2 physical mode with mass parameter $M= \pm i$ on $dS_{4}$, then the resulting eigenmode is a physical mode with the same spin$(4)$ content but with mass parameter $M = \mp i$. Moreover, the matrix $\gamma^{5}$ commutes with the Lie-Lorentz derivative~(\ref{Lie_Lorentz}) with respect to any spin$(4,1)$ Killing vector.}. Thus, below we can just let $M=i$. There are two `chiral' UIR's of spin$(4,1)$ that correspond to the strictly massless spin-3/2 field on $dS_{4}$: one UIR for the helicity $+3/2$ and one UIR for the helicity $-3/2$. The physical modes (i.e. the type-II modes) with helicity $\pm 3/2$ have the following spin$(4)$ content: $\vec{f}^{\pm}_{1} =(\ell+1/2, \pm 
 3/2)$ with $\ell \geq 1$. Let $\vec{F}=(F_{0},F_{1})$ be the spin$(4,1)$ representation formed by the physical modes with helicity $+3/2$. The branching rules~(\ref{branching_rules_spin(2p,1)->spin(2p)}) give $F_{1}=3/2$. Then, by comparing the field-theoretic expression~(\ref{casimir_spin3/2_str_massless}) for the quadratic Casimir with the UIR expression~(\ref{Casimir_Spin(2p,1)}), we find that the physical modes with helicity $+3/2$ form the Discrete Series UIR $D^{+}(\vec{F})=D^{+}(-1/2,3/2)$ [eq.~(\ref{condition_for_discrete_series_+})]. Similarly, we find that the physical modes with helicity $-3/2$ form the Discrete Series UIR $D^{-}(\vec{F})=D^{-}(-1/2,3/2)$ [eq.~(\ref{condition_for_discrete_series_-})]. Thus, the strictly massless spin-3/2 field on $dS_{4}$ corresponds to the direct sum of Discrete Series UIR's $D^{+}(-1/2,3/2) \bigoplus D^{-}(-1/2,3/2)$\footnote{The strictly/partially massless totally symmetric tensors of spin $s=r$ and depth $\uptau = 1,...,r$ on $dS_{4}$ also form a direct sum of Discrete Series UIR's corresponding to $D^{+}(r-\uptau-1,r) \bigoplus D^{-}(r-\uptau-1,r)$~\cite{Yale_Thesis, HiguchiLinearised}.}.
More details can be found in the dictionary in Section~\ref{Sec_Dictionary}.
\section{Dictionary between (symmetric) tensor-spinor fields on \texorpdfstring{$dS_{D}$}{dSD} and UIR's of spin\texorpdfstring{$(D,1)$}{d,1} for \texorpdfstring{$D \geq 3$}{d>=3}} \label{Sec_Dictionary}
Here we present a `field theory - UIR's dictionary' based on our analysis for the spin-3/2 and spin-5/2 eigenmodes satisfying eq.~(\ref{Dirac_eqn_fermion_dS}) on $dS_{D}$. This dictionary relies on the classification of the UIR's under the decomposition spin$(D,1) \supset$ spin$(D)$ given in Section~\ref{Sec_Classification_UIRs} and it was constructed by taking advantage of both:
\begin{itemize}
    \item The values for the spin$(D,1)$ quadratic Casimir corresponding to the eigenmodes [eqs.~(\ref{quadratic_Casimir_Spin(N,1)_analcont_STSSH_spin3/2}), (\ref{quadratic_Casimir_Spin(N,1)_analcont_STSSH_spin5/2}) and~(\ref{casimir_spin3/2_str_massless})-(\ref{casimir_spin5/2_part_massless})].

    \item The spin$(D)$ content of the eigenmodes (see Section~\ref{Section_dS_eigenmodes}, Tables 1 and 2 and Remarks 6.1-6.3).
\end{itemize}
Although until now we have mainly discussed the spin-3/2 and spin-5/2 fields, our analysis and the classification of the UIR's in Section~\ref{Sec_Classification_UIRs} allow us to propose a dictionary for totally symmetric TT tensor-spinors~(\ref{Dirac_eqn_fermion_dS}) with mass parameter $M$ and any half-odd-integer spin $s=r+1/2 \geq  1/2$ on $dS_{D}$ ($D \geq 3$). However, we note that we have not performed an eigenmode analysis for the fields with half-odd-integer spin $s \geq 7/2$ yet, but this is something that we leave for future work. In our dictionary, we give the explicit values for all representation labels concerning the UIR's under the decomposition~spin$(D,1) \supset$ spin$(D)$, and we also translate our results in the representation-theoretic language used in the CFT literature~\cite{Mixed_Symmetry_dS}. While reading the following dictionary, one should recall that the spin$(D)$ content is described by the highest weights of the rank-$\tilde{r}$ TT tensor-spinor eigenmodes~(\ref{tensor-spinoreigen_SD-1}) on $S^{D-1}$: $\vec{f}_{\tilde{r}} = (\ell +1/2, \tilde{r} +1/2, 1/2,...,1/2)$ for odd $D$ [eq.~(\ref{highest_weight_odd_spin(D)_tensor-spinors})] and $\vec{f}^{\pm}_{\tilde{r}} = (\ell +1/2, \tilde{r} +1/2, 1/2,...,1/2,\pm 1/2)$ for even $D$ [eqs.~(\ref{highest_weight_even+_spin(D)_tensor-spinors}) and (\ref{highest_weight_even-_spin(D)_tensor-spinors})] - recall also Remarks 6.1-6.3.

$$\centering \textbf{Dictionary for fields with half-odd-integer spin} ~\bm{s=r+1/2 \geq 1/2}~\textbf{for}~\bm{D \geq 3} $$

\noindent $\bullet$ \textbf{Real $\bm{M \neq 0}$ for all $\bm{D \geq 3}$:} Principal Series UIR's.\\
(This case corresponds to the Principal Series with $so(1,1)$ weight $\Delta_{c}=\tfrac{D-1}{2}-iM$ in Ref.~\cite{Mixed_Symmetry_dS}.)
The representation labels are $\vec{F} = (-\tfrac{D-1}{2} -iM , r+\tfrac{1}{2},\tfrac{1}{2},...,\tfrac{1}{2})$, while for $D \in \{ 3, 4 \}$ we have $\vec{F} = (-\tfrac{D-1}{2} -iM , r+\tfrac{1}{2})$. The {spin${(D)}$ content} corresponds to the highest weights: $\vec{f}_{\tilde{r}}$ (for odd $D \geq 3$) and $\vec{f}^{\pm}_{\tilde{r}}$ (for even $D \geq 4$) with $\ell \geq r \geq \tilde{r} \geq 0$.
For even $D \geq 4$, the eigenmodes with opposite values for their mass parameters form equivalent representations.\\

\noindent $\bullet$ {\textbf{$\bm{M = 0}$ for odd $\bm{D \geq 3}$:}} Principal Series UIR's.\\
(This case corresponds to the Principal Series with $\Delta_{c}=\tfrac{D-1}{2}$ in Ref.~\cite{Mixed_Symmetry_dS}.)
For $D>3$ the {representation labels} are $\vec{F} = (-\tfrac{D-1}{2} , r+\tfrac{1}{2},\tfrac{1}{2},...,\tfrac{1}{2})$, while for $D = 3$ we have $\vec{F} = (-1 , r+\tfrac{1}{2})$. The {spin${(D)}$ content corresponds to the highest weights} $\vec{f}_{\tilde{r}}$ with $\ell \geq r \geq \tilde{r} \geq 0$.\\

\noindent $\bullet$ {\textbf{$\bm{M = 0}$ for even $\bm{D \geq 4}$}:} Direct sum of two Discrete Series UIR's $D^{+}(\vec{F}) \bigoplus D^{-}(\vec{F})$.\\ (In Ref.~\cite{Mixed_Symmetry_dS}, this case corresponds to a direct sum of two Principal Series UIR's with $\Delta_{c}=\tfrac{D-1}{2}$ that are related to each other by space reflection.)
The eigenmodes with spin$(D)$ content $\vec{f}^{+}_{\tilde{r}}$ (where $\ell \geq r \geq \tilde{r} \geq 0$) form the Discrete Series UIR $D^{+} (-\tfrac{D-1}{2} , r+\tfrac{1}{2},\tfrac{1}{2},...,\tfrac{1}{2})$ for $D >4$ and the UIR $D^{+}(-\tfrac{3}{2} , r+\tfrac{1}{2})$ for $D=4$. The eigenmodes with spin$(D)$ content $\vec{f}^{-}_{\tilde{r}}$ (where $\ell \geq r \geq \tilde{r} \geq 0$) form the Discrete Series UIR $D^{-} (-\tfrac{D-1}{2} , r+\tfrac{1}{2},\tfrac{1}{2},...,\tfrac{1}{2})$ for $D >4$ and $D^{-}(-\tfrac{3}{2} , r+\tfrac{1}{2})$ for $D=4$. The eigenmodes that form the UIR $D^{+}$ and the ones forming $D^{-}$ belong to different eigenspaces of the matrix $\gamma^{D+1}$ [eq.~(\ref{gamma(D+1)})].\\

\noindent $\bullet$ {\textbf{Strictly/partially massless fields of depth $\bm{\uptau=1,...,r}$ with $\bm{s \geq 3/2}$ for $\bm{D \neq 4}$}:} Non-unitary.\\

\noindent $\bullet$ {\textbf{Strictly/partially massless fields of depth $\bm{\uptau=1,...,r}$ with $\bm{s \geq 3/2}$ for $\bm{D = 4}$}:} Direct sum of two Discrete Series UIR's of spin$(4,1)$, $D^{+}(\vec{F}) \bigoplus D^{-}(\vec{F})$. \\
(In Ref.~\cite{Mixed_Symmetry_dS}, this case corresponds to a direct sum of two Discrete Series UIR's with $\Delta_{c}=\tfrac{5}{2}+r-\uptau$ that are related to each other by space reflection.)
The physical modes with spin$(D)$ content $\vec{f}^{+}_{\tilde{r}}= (\ell+\tfrac{1}{2}, \tilde{r}+\tfrac{1}{2})$ (where $\ell\geq r \geq \tilde{r} \geq r -\uptau +1$) form the Discrete Series UIR $D^{+}(r-\uptau-\tfrac{1}{2} , r+\tfrac{1}{2})$. The physical modes with spin$(D)$ content $\vec{f}^{-}_{\tilde{r}}=(\ell+\tfrac{1}{2}, -\tilde{r}-\tfrac{1}{2})$ (where $\ell \geq r \geq \tilde{r} \geq r -\uptau+1$) form the Discrete Series UIR $D^{-}(r - \uptau-\tfrac{1}{2} , r+\tfrac{1}{2})$. In particular, the UIR $D^{\pm}(r-\uptau-\tfrac{1}{2} , r+\tfrac{1}{2})$ corresponds to the depth-$\uptau$ field with propagating helicities $(\pm s, \pm(s-1), ..., \pm( s-\uptau+1))$. In the strictly massless case ($\uptau =1$), the UIR $D^{+}(r-\tfrac{3}{2} , r+\tfrac{1}{2})$ corresponds to the single helicity $s$, while $D^{-}(r-\tfrac{3}{2} , r+\tfrac{1}{2})$ corresponds to the single helicity $-s$.
No physical (or pure gauge~(\ref{PG_TYPEI_spin3/2})) mode is an eigenfunction of the matrix $\gamma^{5}$ [eq.~(\ref{gamma(D+1)})].
 \section{Summary and discussions}
 In the present paper, we demonstrated that four-dimensional dS space plays a distinguished role in the unitarity of the strictly and partially massless (symmetric) tensor-spinor fields of spin $s = 3/2, 5/2$. In particular, the strictly massless spin-3/2 field, as well as the strictly and partially massless spin-5/2 fields on $dS_{D}$, are not unitary unless $D=4$. The explanation relies on the representation theory of spin$(D,1)$, where the latter does not allow strictly/partially massless UIR's for (symmetric) tensor-spinors unless $D=4$. This is a remarkable feature of dS field theory, while it is also very interesting that the dimensionality that plays a special representation-theoretic role matches the dimensionality of our physical Universe. We also expect that this result should hold for all totally symmetric tensor-spinors with spin $s \geq 7/2$, while this expectation of ours is justified by the classification of the spin$(D,1)$ UIR's. A technical explanation of our results in terms of the (non-)existence of positive-definite dS scalar products for the spin-$3/2$ and spin-$5/2$ eigenmodes has been given in Refs.~\cite{Letsios_in_progress, Letsios_arxiv_long}.
 
In Section~\ref{Sec_Dictionary}, we presented a dictionary between (totally symmetric) half-odd-integer-spin fields on $dS_{D}$ and UIR's of spin$(D,1)$ ($D \geq 3$). The validity of our dictionary for the spin-3/2 and spin-5/2 fields was demonstrated in this paper. Our dictionary for the cases with half-odd-integer spin $s \geq 7/2$ is a `suggestion' that is motivated by the classification of the UIR's and can be confirmed by performing an eigenmode analysis for half-odd-integer spins $s \geq 7/2$. This is something that we leave for future work.

In the present paper, `unitarity' of a field theory does not just refer to the positivity of the norm in the Hilbert space. In this paper, unitarity in the one-particle Hilbert space means that: a positive-definite scalar product for the eigenmodes exists that is invariant under spin$(D,1)$. If and only if these conditions are satisfied then the space of eigenmodes can be identified with the representation space of a unitary representation of spin$(D,1)$. For example, consider the strictly massless spin-3/2 field on $dS_{4}$ satisfying the onshell conditions
\begin{align}\label{Dirac_spin3/2}
   &\left( \slashed{\nabla}\pm i\right)\Psi_{\mu}=0\\
   & \nabla^{\alpha}\Psi_{\alpha}=0, \hspace{4mm}  \gamma^{\alpha}\Psi_{\alpha }=0. \label{TT_conditions_fermions_dS_3/2}
\end{align}
The physical eigenmodes of this theory are given by eqs.~(\ref{TYPE_II_negative_spin_3/2_anal_cont}) and~(\ref{TYPE_II_positive_spin_3/2_anal_cont}). It is easy to check that the following (Dirac-like) scalar product is positive-definite
\begin{align}\label{conventional_scalar_prod}
    \int_{S^{3}} \sqrt{-g}\, d\bm{\theta_{3}}\, g^{\mu \nu} \Psi^{(1)\dagger}_{\mu}(t, \bm{\theta_{3}})\, \Psi^{(2)}_{\nu}(t, \bm{\theta_{3}})
\end{align}
for any two physical modes $\Psi^{(1)}_{\mu}$ and $\Psi^{(2)}_{\nu}$, where $g$ is the determinant of the $dS_{4}$ metric, while $d \bm{\theta_{3}}$ stands for $d \theta_{3}  d \theta_{2}  d\theta_{1}$. This is the scalar product for the one-particle Hilbert space that was implicitly used in order to check the positivity property of the equal time anti-commutators in Ref.~\cite{Deser_Waldron_phases}. However, while the positivity of the norm with respect to the scalar product~(\ref{conventional_scalar_prod}) is clearly necessary, it is not sufficient for representation-theoretic unitarity. In particular, it is straightforward to check that the scalar product~(\ref{conventional_scalar_prod}) is neither conserved nor dS invariant~\cite{Letsios_in_progress, Letsios_arxiv_long}. The reason is that the conventional (Dirac-like) vector current
\begin{equation}\label{Dirac_current}
     J^{\mu}= -\,{\Psi}^{(1)\dagger}_{\nu} \gamma^{0} \gamma^{\mu} \Psi^{(2)\nu}
    \end{equation}
    is not covariantly conserved because of the imaginary mass parameter in eq.~(\ref{Dirac_spin3/2}). Thus, we cannot use the scalar product~(\ref{conventional_scalar_prod}) in order to check the unitarity of the spin$(4,1)$ representation formed by the physical modes. On the other hand,
    the (axial) vector current
\begin{equation}\label{axial_current}
     J_{\text{ax}}^{\mu}= -\,{\Psi}^{(1)\dagger}_{\nu} \gamma^{0} \gamma^{\mu} \gamma^{5} \Psi^{(2)\nu}
    \end{equation}
    is covariantly conserved, giving rise to the time-independent and dS invariant scalar product~\cite{Letsios_in_progress, Letsios_arxiv_long}
\begin{align}\label{axial_scalar_prod}
\int_{S^{3}} \sqrt{-g}\, d\bm{\theta_{3}}J_{\text{ax}}^{0}=     \int_{S^{3}} \sqrt{-g}\, d\bm{\theta_{3}}\, g^{\mu \nu} \Psi^{(1)\dagger}_{\mu}(t, \bm{\theta_{3}})\,\gamma^{5}\, \Psi^{(2)}_{\nu}(t, \bm{\theta_{3}}).
\end{align}
This scalar product is a good choice in order to study the unitarity of the corresponding spin$(4,1)$ representation for the reasons mentioned above. In particular, the physical modes~(\ref{TYPE_II_negative_spin_3/2_anal_cont}) form the Discrete Series UIR $D^{-}(-\tfrac{1}{2} , \tfrac{3}{2})$ with the positive-definite scalar product~(\ref{axial_scalar_prod}), while the physical modes (\ref{TYPE_II_positive_spin_3/2_anal_cont}) form the Discrete Series UIR $D^{+}(-\tfrac{1}{2} , \tfrac{3}{2})$ with positive-definite scalar product given by the negative of eq.~(\ref{axial_scalar_prod}). The pure gauge modes~(\ref{PG_TYPEI_spin3/2}) have zero norm with respect to the scalar product~(\ref{axial_scalar_prod}) and they are orthogonal to all physical modes. For more details concerning the eigenmodes see Refs.~\cite{Letsios_in_progress, Letsios_arxiv_long}.

The fermionic strictly and partially massless tunings~(\ref{values_mass_parameter_masslessness_fermion}) were found in Ref.~\cite{Deser_Waldron_ArbitrarySR}, but the non-unitarity of the corresponding theories for $D \neq 4$ could not be revealed with the methods used in this reference.


\acknowledgments

 The author is grateful to Atsushi Higuchi for guidance, suggestions, and encouragement, as well as for invaluable discussions and ideas that inspired the majority of the material presented in this paper and helpful comments on earlier versions of this paper. Also, it is a pleasure to thank Stanley Deser for communications and Andrew Waldron for useful discussions. The author would also like to thank the referee for their useful comments and suggestions. The author also thanks Xavier Bekaert, Nicolas Boulanger, Thomas Basile, Charis Anastopoulos, Lasse Schmieding, Nikolaos Koutsonikos-Kouloumpis, and F. F. John for useful discussions. He also thanks Zoi Taglee for insightful and inspiring discussions. This work was supported by a studentship from the Department of Mathematics, University of York.

\appendix 

\section{The only totally symmetric TT tensor-spinor eigenmodes of the Dirac operator that exist on \texorpdfstring{$S^{2}$}{S2} are the spinor eigenmodes}\label{Appendix_tensor-spinors_twosphere}
The spinor eigenmodes of the Dirac operator on $S^{2}$ (as well as on spheres of any dimension) have been constructed in Ref.~\cite{Camporesi}.
In Ref.~\cite{CHH}, the TT vector-spinor eigenmodes of the Dirac operator on $S^{d}$ ($d \geq 3$) were obtained, while it was found that there are no TT vector-spinor eigenmodes on $S^{2}$. In Refs.~\cite{Letsios_in_progress, Letsios_arxiv_long}, the author has constructed the rank-2 symmetric TT tensor-spinor eigenmodes of the Dirac operator on $S^{d}$ ($d \geq 3$) and he also found that such eigenmodes do not exist on $S^{2}$.
In this Appendix, we will show that there are no totally symmetric TT tensor-spinor eigenmodes $\tilde{\psi}_{\tilde{\mu}_{1}...\tilde{\mu}_{\tilde{r}}}$ of rank $\tilde{r} \geq 2$ on $S^{2}$. Our proof will closely follow the analogous proof for totally symmetric tensors of rank $\tilde{r} \geq 2$ on $S^{2}$ in Ref.~\cite{STSHS}. For convenience, we will drop the tildes from the tensor indices and, thus, our tensor-spinor of rank $\tilde{r} \geq 2$ on $S^{2}$ will be denoted as $\tilde{\psi}_{{\mu}_{1}...{\mu}_{\tilde{r}}}$.

For later convenience, note that the Riemann tensor on $S^{2}$ is \begin{align}\label{Riemann_tens}
    \tilde{R}_{{\mu} {\nu} {\kappa} {\lambda}}=\tilde{g}_{{\mu} {\kappa}} \tilde{g}_{{\nu} {\lambda}}-\tilde{g}_{{\nu}{\kappa}} \tilde{g}_{{\mu} {\lambda}},
\end{align}
where $\tilde{g}_{{\mu}  {\nu}}$ is the metric tensor on $S^{2}$. The commutator of covariant derivatives acting on a vector-spinor on $S^{2}$ is given by
\begin{align}
    [\tilde{\nabla}_{\mu}, \tilde{\nabla}_{\nu}]\tilde{\psi}_{\alpha}&=\frac{1}{4}\tilde{R}_{\mu \nu \kappa \lambda} \tilde{\gamma}^{\kappa}\tilde{\gamma}^{\lambda}\tilde{\psi}_{\alpha}+\tilde{R}^{\lambda}_{\hspace{1mm}\alpha \nu \mu}\tilde{\psi}_{\lambda}\\
    &=\frac{1}{2}(\tilde{\gamma}_{\mu}\tilde{\gamma}_{\nu}-\tilde{g}_{\mu \nu})\tilde{\psi}_{\alpha}+2\tilde{g}_{\alpha[\mu}\tilde{\psi}_{\nu]},\label{commutr_deriv_vecspinor}
\end{align}
where $\tilde{\gamma}_{\mu}$ are the gamma matrices on $S^{2}$. The expressions for the commutators of covariant derivatives for tensor-spinors of higher rank are straightforward generalisations of eq.~(\ref{commutr_deriv_vecspinor}). Also, let $\tilde{\epsilon}_{\mu \nu}$ be the anti-symmetric tensor on $S^{2}$. In the coordinate system~(\ref{line_element_sphere_inductive}), $\tilde{\epsilon}_{\mu \nu}$ is defined by
\begin{align}
&  \tilde{\epsilon}_{\theta_{1}  \theta_{1}} =  \tilde{\epsilon}_{\theta_{2} \theta_{2}} =0\nonumber \\
 &   \tilde{\epsilon}_{\theta_{2}  \theta_{1}}= - \tilde{\epsilon}_{\theta_{1}  \theta_{2}}= \sin{\theta_{2}},
\end{align}
where $\tilde{\nabla}_{\alpha} \tilde{\epsilon}_{\mu \nu} = 0$.
Now, let us define
\begin{align}
   \tilde{\nabla}_{ [ \theta_{1}} \tilde{\psi}_{{\theta_{2}}] {\mu}_{2}...{\mu}_{\tilde{r}}} = \tilde{\epsilon}_{\theta_{1}   \theta_{2}} A_{\mu_{2}...\mu_{\tilde{r}}},
\end{align}
where $A_{\mu_{2}...\mu_{\tilde{r}}}$ is a totally symmetric tensor-spinor of rank $\tilde{r}-1$ on $S^{2}$.
Then
\begin{align}\label{anti-sym-tensorspinor}
   \tilde{\nabla}_{ [ {\mu}} \tilde{\psi}_{{\nu}] {\mu}_{2}...{\mu}_{\tilde{r}}} = \tilde{\epsilon}_{\mu \nu } A_{\mu_{2}...\mu_{\tilde{r}}}.
\end{align}
 By taking the trace of eq.~(\ref{anti-sym-tensorspinor}) with respect to the indices $\nu$ and $\mu_{2}$, and by using the fact that $\tilde{\psi}_{{\nu} {\mu}_{2}...{\mu}_{\tilde{r}}}$ is traceless and divergence-free, we find $A_{\mu_{2}...\mu_{\tilde{r}}} = 0$. In other words,
\begin{align}
   \tilde{\nabla}_{ [ {\mu}} \tilde{\psi}_{{\nu}] {\mu}_{2}...{\mu}_{\tilde{r}}}=0.
\end{align}
By taking the divergence of this equation with respect to the index $\mu$, and making use of eq.~(\ref{commutr_deriv_vecspinor}), we find
\begin{align}
   \tilde{\nabla}_{{\mu}}\tilde{\nabla}^{\mu} \tilde{\psi}_{{\nu} {\mu}_{2}...{\mu}_{\tilde{r}}}= \left(\tilde{r} + \frac{1}{2}\right)\tilde{\psi}_{{\nu} {\mu}_{2}...{\mu}_{\tilde{r}}}.
\end{align}
However, as is well-known, $\tilde{\nabla}_{{\mu}}\tilde{\nabla}^{\mu}$ is negative-definite on compact manifolds. Thus, $\tilde{\psi}_{{\nu} {\mu}_{2}...{\mu}_{\tilde{r}}}$ must be identically zero.

\providecommand{\noopsort}[1]{}\providecommand{\singleletter}[1]{#1}%


\begin{thebibliography}{45}%
\makeatletter
\providecommand \@ifxundefined [1]{%
 \@ifx{#1\undefined}
}%
\providecommand \@ifnum [1]{%
 \ifnum #1\expandafter \@firstoftwo
 \else \expandafter \@secondoftwo
 \fi
}%
\providecommand \@ifx [1]{%
 \ifx #1\expandafter \@firstoftwo
 \else \expandafter \@secondoftwo
 \fi
}%
\providecommand \natexlab [1]{#1}%
\providecommand \enquote  [1]{``#1''}%
\providecommand \bibnamefont  [1]{#1}%
\providecommand \bibfnamefont [1]{#1}%
\providecommand \citenamefont [1]{#1}%
\providecommand \href@noop [0]{\@secondoftwo}%
\providecommand \href [0]{\begingroup \@sanitize@url \@href}%
\providecommand \@href[1]{\@@startlink{#1}\@@href}%
\providecommand \@@href[1]{\endgroup#1\@@endlink}%
\providecommand \@sanitize@url [0]{\catcode `\\12\catcode `\$12\catcode
  `\&12\catcode `\#12\catcode `\^12\catcode `\_12\catcode `\%12\relax}%
\providecommand \@@startlink[1]{}%
\providecommand \@@endlink[0]{}%
\providecommand \url  [0]{\begingroup\@sanitize@url \@url }%
\providecommand \@url [1]{\endgroup\@href {#1}{\urlprefix }}%
\providecommand \urlprefix  [0]{URL }%
\providecommand \Eprint [0]{\href }%
\providecommand \doibase [0]{https://doi.org/}%
\providecommand \selectlanguage [0]{\@gobble}%
\providecommand \bibinfo  [0]{\@secondoftwo}%
\providecommand \bibfield  [0]{\@secondoftwo}%
\providecommand \translation [1]{[#1]}%
\providecommand \BibitemOpen [0]{}%
\providecommand \bibitemStop [0]{}%
\providecommand \bibitemNoStop [0]{.\EOS\space}%
\providecommand \EOS [0]{\spacefactor3000\relax}%
\providecommand \BibitemShut  [1]{\csname bibitem#1\endcsname}%
\let\auto@bib@innerbib\@empty
\bibitem [{\citenamefont {{SUPERNOVA COSMOLOGY PROJECT
  collaboration}}(1999)}]{Perlmutter_1999}%
  \BibitemOpen
  \bibfield  {author} {\bibinfo {author} {\bibnamefont {{SUPERNOVA COSMOLOGY
  PROJECT collaboration}}},\ }\bibfield  {title} {\bibinfo {title}
  {Measurements of ${\Omega}$ and ${\Lambda}$ from 42 {H}igh-{R}edshift
  {S}upernovae},\ }\href {https://doi.org/10.1086/307221} {\bibfield  {journal}
  {\bibinfo  {journal} {Astrophys. J.}\ }\textbf {\bibinfo {volume} {517}},\
  \bibinfo {pages} {565} (\bibinfo {year} {1999})}\BibitemShut {NoStop}%
\bibitem [{\citenamefont {{SDSS collaboration}}(2010)}]{SloanDigitalSky}%
  \BibitemOpen
  \bibfield  {author} {\bibinfo {author} {\bibnamefont {{SDSS
  collaboration}}},\ }\bibfield  {title} {\bibinfo {title} {{Baryon acoustic
  oscillations in the Sloan Digital Sky Survey Data Release 7 Galaxy Sample}},\
  }\href {https://doi.org/10.1111/j.1365-2966.2009.15812.x} {\bibfield
  {journal} {\bibinfo  {journal} {Mon. Not. Roy. Astron. Soc.}\ }\textbf
  {\bibinfo {volume} {401}},\ \bibinfo {pages} {2148} (\bibinfo {year}
  {2010})},\ \Eprint
  {https://arxiv.org/abs/https://academic.oup.com/mnras/article-pdf/401/4/2148/3901461/mnras0401-2148.pdf}
  {https://academic.oup.com/mnras/article-pdf/401/4/2148/3901461/mnras0401-2148.pdf}
  \BibitemShut {NoStop}%
\bibitem [{\citenamefont {{PLANCK collaboration}}(2020)}]{PlanckCollab}%
  \BibitemOpen
  \bibfield  {author} {\bibinfo {author} {\bibnamefont {{PLANCK
  collaboration}}},\ }\bibfield  {title} {\bibinfo {title} {Planck 2018 results
  - {VI.} {C}osmological parameters},\ }\href
  {https://doi.org/10.1051/0004-6361/201833910} {\bibfield  {journal} {\bibinfo
   {journal} {Astron. Astrophys.}\ }\textbf {\bibinfo {volume} {641}},\
  \bibinfo {pages} {A6} (\bibinfo {year} {2020})}\BibitemShut {NoStop}%
\bibitem [{\citenamefont {Hawking}\ and\ \citenamefont
  {Ellis}(1973)}]{hawking_ellis_1973}%
  \BibitemOpen
  \bibfield  {author} {\bibinfo {author} {\bibfnamefont {S.~W.}\ \bibnamefont
  {Hawking}}\ and\ \bibinfo {author} {\bibfnamefont {G.~F.~R.}\ \bibnamefont
  {Ellis}},\ }\href {https://doi.org/10.1017/CBO9780511524646} {\emph {\bibinfo
  {title} {The Large Scale Structure of Space-Time}}},\ Cambridge Monographs on
  Mathematical Physics\ (\bibinfo  {publisher} {Cambridge University Press},\
  \bibinfo {year} {1973})\BibitemShut {NoStop}%
\bibitem [{\citenamefont {Tung}(1985)}]{Tung}%
  \BibitemOpen
  \bibfield  {author} {\bibinfo {author} {\bibfnamefont {W.-K.}\ \bibnamefont
  {Tung}},\ }\href {https://doi.org/10.1142/0097} {\emph {\bibinfo {title}
  {Group Theory in Physics: An Introduction To Symmetry Principles, Group
  Representations, And Special Functions In Classical And Quantum Physics}}}\
  (\bibinfo  {publisher} {World Scientific},\ \bibinfo {year} {1985})\ \Eprint
  {https://arxiv.org/abs/https://www.worldscientific.com/doi/pdf/10.1142/0097}
  {https://www.worldscientific.com/doi/pdf/10.1142/0097} \BibitemShut {NoStop}%
\bibitem [{\citenamefont {Deser}\ and\ \citenamefont
  {Waldron}(2001{\natexlab{a}})}]{Deser_Waldron_null_propagation}%
  \BibitemOpen
  \bibfield  {author} {\bibinfo {author} {\bibfnamefont {S.}~\bibnamefont
  {Deser}}\ and\ \bibinfo {author} {\bibfnamefont {A.}~\bibnamefont
  {Waldron}},\ }\bibfield  {title} {\bibinfo {title} {{Null propagation of
  partially massless higher spins in (A)dS and cosmological constant
  speculations}},\ }\href {https://doi.org/10.1016/S0370-2693(01)00756-0}
  {\bibfield  {journal} {\bibinfo  {journal} {Phys. Lett. B}\ }\textbf
  {\bibinfo {volume} {513}},\ \bibinfo {pages} {137} (\bibinfo {year}
  {2001}{\natexlab{a}})},\ \Eprint {https://arxiv.org/abs/hep-th/0105181}
  {arXiv:hep-th/0105181} \BibitemShut {NoStop}%
\bibitem [{\citenamefont {Deser}\ and\ \citenamefont
  {Waldron}(2001{\natexlab{b}})}]{Deser_Waldron_stability_of_massive_cosm}%
  \BibitemOpen
  \bibfield  {author} {\bibinfo {author} {\bibfnamefont {S.}~\bibnamefont
  {Deser}}\ and\ \bibinfo {author} {\bibfnamefont {A.}~\bibnamefont
  {Waldron}},\ }\bibfield  {title} {\bibinfo {title} {{Stability of massive
  cosmological gravitons}},\ }\href
  {https://doi.org/10.1016/S0370-2693(01)00523-8} {\bibfield  {journal}
  {\bibinfo  {journal} {Phys. Lett. B}\ }\textbf {\bibinfo {volume} {508}},\
  \bibinfo {pages} {347} (\bibinfo {year} {2001}{\natexlab{b}})},\ \Eprint
  {https://arxiv.org/abs/hep-th/0103255} {arXiv:hep-th/0103255} \BibitemShut
  {NoStop}%
\bibitem [{\citenamefont {Deser}\ and\ \citenamefont
  {Waldron}(2001{\natexlab{c}})}]{Deser_Waldron_phases}%
  \BibitemOpen
  \bibfield  {author} {\bibinfo {author} {\bibfnamefont {S.}~\bibnamefont
  {Deser}}\ and\ \bibinfo {author} {\bibfnamefont {A.}~\bibnamefont
  {Waldron}},\ }\bibfield  {title} {\bibinfo {title} {{Gauge invariances and
  phases of massive higher spins in (A)dS}},\ }\href
  {https://doi.org/10.1103/PhysRevLett.87.031601} {\bibfield  {journal}
  {\bibinfo  {journal} {Phys. Rev. Lett.}\ }\textbf {\bibinfo {volume} {87}},\
  \bibinfo {pages} {031601} (\bibinfo {year} {2001}{\natexlab{c}})},\ \Eprint
  {https://arxiv.org/abs/hep-th/0102166} {arXiv:hep-th/0102166} \BibitemShut
  {NoStop}%
\bibitem [{\citenamefont {Deser}\ and\ \citenamefont
  {Waldron}(2001{\natexlab{d}})}]{Deser_Waldron_partial_masslessness}%
  \BibitemOpen
  \bibfield  {author} {\bibinfo {author} {\bibfnamefont {S.}~\bibnamefont
  {Deser}}\ and\ \bibinfo {author} {\bibfnamefont {A.}~\bibnamefont
  {Waldron}},\ }\bibfield  {title} {\bibinfo {title} {Partial masslessness of
  higher spins in ({A})d{S}},\ }\href@noop {} {\bibfield  {journal} {\bibinfo
  {journal} {Nuclear Physics}\ }\textbf {\bibinfo {volume} {607}},\ \bibinfo
  {pages} {577} (\bibinfo {year} {2001}{\natexlab{d}})}\BibitemShut {NoStop}%
\bibitem [{\citenamefont {Deser}\ and\ \citenamefont
  {A.Waldron}(2004)}]{Deser_Waldron_Conformal}%
  \BibitemOpen
  \bibfield  {author} {\bibinfo {author} {\bibfnamefont {S.}~\bibnamefont
  {Deser}}\ and\ \bibinfo {author} {\bibnamefont {A.Waldron}},\ }\bibfield
  {title} {\bibinfo {title} {Conformal invariance of partially massless higher
  spins},\ }\href@noop {} {\bibfield  {journal} {\bibinfo  {journal} {Physics
  Letters B}\ }\textbf {\bibinfo {volume} {603}},\ \bibinfo {pages} {30}
  (\bibinfo {year} {2004})}\BibitemShut {NoStop}
\bibitem [{\citenamefont {Deser}\ and\ \citenamefont
  {Nepomechie}(1983)}]{DESER_NEPOM_1}%
  \BibitemOpen
  \bibfield  {author} {\bibinfo {author} {\bibfnamefont {S.}~\bibnamefont
  {Deser}}\ and\ \bibinfo {author} {\bibfnamefont {R.~I.}\ \bibnamefont
  {Nepomechie}},\ }\bibfield  {title} {\bibinfo {title} {Anomalous propagation
  of gauge fields in conformally flat spaces},\ }\href
  {https://doi.org/https://doi.org/10.1016/0370-2693(83)90317-9} {\bibfield
  {journal} {\bibinfo  {journal} {Physics Letters B}\ }\textbf {\bibinfo
  {volume} {132}},\ \bibinfo {pages} {321} (\bibinfo {year}
  {1983})}\BibitemShut {NoStop}%
\bibitem [{\citenamefont {Deser}\ and\ \citenamefont
  {Nepomechie}(1984)}]{DESER_NEPOM_2}%
  \BibitemOpen
  \bibfield  {author} {\bibinfo {author} {\bibfnamefont {S.}~\bibnamefont
  {Deser}}\ and\ \bibinfo {author} {\bibfnamefont {R.~I.}\ \bibnamefont
  {Nepomechie}},\ }\bibfield  {title} {\bibinfo {title} {Gauge invariance
  versus masslessness in de sitter spaces},\ }\href
  {https://doi.org/https://doi.org/10.1016/0003-4916(84)90156-8} {\bibfield
  {journal} {\bibinfo  {journal} {Annals of Physics}\ }\textbf {\bibinfo
  {volume} {154}},\ \bibinfo {pages} {396} (\bibinfo {year}
  {1984})}\BibitemShut {NoStop}
  \bibitem [{\citenamefont {Lust}(1983)}]{Lust}%
  \BibitemOpen
  \bibfield  {author} {\bibinfo {author} {\bibfnamefont {D.}~\bibnamefont
  {L{\"u}st} }\ and\ \bibinfo {author} {\bibfnamefont {E.}\ \bibnamefont
  {Palti}},\ }\bibfield  {title} {\bibinfo {title} {A note on string excitations and the Higuchi bound,\ } }\href {https://doi.org/10.1016/j.physletb.2019.135067}
  {\bibfield  {journal} {\bibinfo  {journal} {Physics Letters B}\ }\textbf
  {\bibinfo {volume} {799}},\ \bibinfo {pages} {135067} (\bibinfo {year}
  {2019})}\BibitemShut {NoStop} 
\bibitem [{\citenamefont {Higuchi}(1987{\natexlab{a}})}]{STSHS}%
  \BibitemOpen
  \bibfield  {author} {\bibinfo {author} {\bibfnamefont {A.}~\bibnamefont
  {Higuchi}},\ }\bibfield  {title} {\bibinfo {title} {Symmetric tensor
  spherical harmonics on the {$N$}-sphere and their application to the de
  {S}itter group {${\rm SO}(N,1)$}},\ }\href {https://doi.org/10.1063/1.527513}
  {\bibfield  {journal} {\bibinfo  {journal} {J. Math. Phys.}\ }\textbf
  {\bibinfo {volume} {28}},\ \bibinfo {pages} {1553} (\bibinfo {year}
  {1987}{\natexlab{a}})}\BibitemShut {NoStop}%
\bibitem [{\citenamefont {Deser}\ and\ \citenamefont
  {Waldron}(2003)}]{Deser_Waldron_ArbitrarySR}%
  \BibitemOpen
  \bibfield  {author} {\bibinfo {author} {\bibfnamefont {S.}~\bibnamefont
  {Deser}}\ and\ \bibinfo {author} {\bibfnamefont {A.}~\bibnamefont
  {Waldron}},\ }\bibfield  {title} {\bibinfo {title} {Arbitrary spin
  representations in de {S}itter from d{S} / {CFT} with applications to d{S}
  supergravity},\ }\href@noop {} {\bibfield  {journal} {\bibinfo  {journal}
  {Nuclear Physics}\ }\textbf {\bibinfo {volume} {662}},\ \bibinfo {pages}
  {379} (\bibinfo {year} {2003})}\BibitemShut {NoStop}%
\bibitem [{\citenamefont {Higuchi}(1987{\natexlab{b}})}]{Yale_Thesis}%
  \BibitemOpen
  \bibfield  {author} {\bibinfo {author} {\bibfnamefont {A.}~\bibnamefont
  {Higuchi}},\ }\bibfield  {title} {\bibinfo {title} {Quantum fields of nonzero
  spin in {D}e {S}itter spacetime},\ }\href@noop {} {\bibfield  {journal}
  {\bibinfo  {journal} {PhD dissertation, Yale University}\ } (\bibinfo {year}
  {1987}{\natexlab{b}})}\BibitemShut {NoStop}%
  \bibitem [{\citenamefont {Basile}\ \emph {et~al.}(2016)\citenamefont {Basile},
  \citenamefont {Bekaert},\ and\ \citenamefont
  {Boulanger}}]{Mixed_Symmetry_dS}%
  \BibitemOpen
  \bibfield  {author} {\bibinfo {author} {\bibfnamefont {T.}~\bibnamefont
  {Basile}}, \bibinfo {author} {\bibfnamefont {X.}~\bibnamefont {Bekaert}},\
  and\ \bibinfo {author} {\bibfnamefont {N.}~\bibnamefont {Boulanger}},\
  }\bibfield  {title} {\bibinfo {title} {Mixed-symmetry fields in de Sitter
  space: a group theoretical glance},\ }\href@noop {} {\bibfield  {journal}
  {\bibinfo  {journal} {Journal of High Energy Physics}\ }\textbf {\bibinfo
  {volume} {2017}},\ \bibinfo {pages} {1} (\bibinfo {year} {2016})}\BibitemShut
  {NoStop}%
\bibitem [{\citenamefont {Chen}\ \emph {et~al.}(2016)\citenamefont {Chen},
  \citenamefont {Cho}, \citenamefont {Cornell},\ and\ \citenamefont
  {Harmsen}}]{CHH}%
  \BibitemOpen
  \bibfield  {author} {\bibinfo {author} {\bibfnamefont {C.-H.}\ \bibnamefont
  {Chen}}, \bibinfo {author} {\bibfnamefont {H.~T.}\ \bibnamefont {Cho}},
  \bibinfo {author} {\bibfnamefont {A.~S.}\ \bibnamefont {Cornell}},\ and\
  \bibinfo {author} {\bibfnamefont {G.}~\bibnamefont {Harmsen}},\ }\bibfield
  {title} {\bibinfo {title} {Spin-3/2 fields in {$D$}-dimensional
  {S}chwarzschild black hole spacetimes},\ }\href
  {https://doi.org/10.1103/PhysRevD.94.044052} {\bibfield  {journal} {\bibinfo
  {journal} {Phys. Rev. D}\ }\textbf {\bibinfo {volume} {94}},\ \bibinfo
  {pages} {044052} (\bibinfo {year} {2016})}\BibitemShut {NoStop}%
\bibitem [{\citenamefont {U.}(1968)}]{Ottoson}%
  \BibitemOpen
  \bibfield  {author} {\bibinfo {author} {\bibfnamefont {U.}~\bibnamefont
  {Ottoson}},\ }\bibfield  {title} {\bibinfo {title} {{A Classification of the
  Unitary Irreducible Representations of $SO_{0}(N,1)$}},\ }\href
  {https://doi.org/10.1007/BF01645858} {\bibfield  {journal} {\bibinfo
  {journal} {Commun. Math. Phys.}\ }\textbf {\bibinfo {volume} {8}},\ \bibinfo
  {pages} {228} (\bibinfo {year} {1968})}\BibitemShut {NoStop}%
\bibitem [{\citenamefont {Schwarz}(1971)}]{Schwarz}%
  \BibitemOpen
  \bibfield  {author} {\bibinfo {author} {\bibfnamefont {F.}~\bibnamefont
  {Schwarz}},\ }\bibfield  {title} {\bibinfo {title} {Unitary {I}rreducible
  {R}epresentations of the {G}roups ${SO}_{0}(n, 1)$},\ }\href
  {https://doi.org/10.1063/1.1665471} {\bibfield  {journal} {\bibinfo
  {journal} {Journal of Mathematical Physics}\ }\textbf {\bibinfo {volume}
  {12}},\ \bibinfo {pages} {131} (\bibinfo {year} {1971})},\ \Eprint
  {https://arxiv.org/abs/https://doi.org/10.1063/1.1665471}
  {https://doi.org/10.1063/1.1665471} \BibitemShut {NoStop}%
\bibitem [{\citenamefont {Camporesi}\ and\ \citenamefont
  {Higuchi}(1996)}]{Camporesi}%
  \BibitemOpen
  \bibfield  {author} {\bibinfo {author} {\bibfnamefont {R.}~\bibnamefont
  {Camporesi}}\ and\ \bibinfo {author} {\bibfnamefont {A.}~\bibnamefont
  {Higuchi}},\ }\bibfield  {title} {\bibinfo {title} {On the eigenfunctions of
  the {D}irac operator on spheres and real hyperbolic spaces},\ }\href
  {https://doi.org/10.1016/0393-0440(95)00042-9} {\bibfield  {journal}
  {\bibinfo  {journal} {J. Geom. Phys.}\ }\textbf {\bibinfo {volume} {20}},\
  \bibinfo {pages} {1} (\bibinfo {year} {1996})}\BibitemShut {NoStop}%
\bibitem [{\citenamefont {A.~Letsios}(2021)}]{Letsios}%
  \BibitemOpen
  \bibfield  {author} {\bibinfo {author} {\bibfnamefont {V.}~\bibnamefont
  {A.~Letsios}},\ }\bibfield  {title} {\bibinfo {title} {The eigenmodes for
  spinor quantum field theory in global de {S}itter spacetime},\ }\href
  {https://doi.org/10.1063/5.0038651} {\bibfield  {journal} {\bibinfo
  {journal} {Journal of Mathematical Physics}\ }\textbf {\bibinfo {volume}
  {62}},\ \bibinfo {pages} {032303} (\bibinfo {year} {2021})},\ \Eprint
  {https://arxiv.org/abs/https://doi.org/10.1063/5.0038651}
  {https://doi.org/10.1063/5.0038651} \BibitemShut {NoStop}
\bibitem [{\citenamefont {Gradshteyn}\ and\ \citenamefont
  {Ryzhik}(2007)}]{gradshteyn2007}%
  \BibitemOpen
  \bibfield  {author} {\bibinfo {author} {\bibfnamefont {I.~S.}\ \bibnamefont
  {Gradshteyn}}\ and\ \bibinfo {author} {\bibfnamefont {I.~M.}\ \bibnamefont
  {Ryzhik}},\ }\href@noop {} {\emph {\bibinfo {title} {Table of integrals,
  series, and products}}},\ \bibinfo {edition} {seventh}\ ed.\ (\bibinfo
  {publisher} {Elsevier/Academic Press, Amsterdam},\ \bibinfo {year} {2007})\
  pp.\ \bibinfo {pages} {xlviii+1171},\ \bibinfo {note} {translated from the
  Russian, Translation edited and with a preface by Alan Jeffrey and Daniel
  Zwillinger, With one CD-ROM (Windows, Macintosh and UNIX)}\BibitemShut
  {NoStop}%
\bibitem [{\citenamefont {Barut}\ and\ \citenamefont
  {Raczka}(1986)}]{barut_group}%
  \BibitemOpen
  \bibfield  {author} {\bibinfo {author} {\bibfnamefont {A.}~\bibnamefont
  {Barut}}\ and\ \bibinfo {author} {\bibfnamefont {R.}~\bibnamefont {Raczka}},\
  }\href {https://doi.org/10.1142/0352} {\emph {\bibinfo {title} {Theory of
  Group Representations and Applications}}}\ (\bibinfo  {publisher} {WORLD
  SCIENTIFIC},\ \bibinfo {year} {1986})\ \Eprint
  {https://arxiv.org/abs/https://www.worldscientific.com/doi/pdf/10.1142/0352}
  {https://www.worldscientific.com/doi/pdf/10.1142/0352} \BibitemShut {NoStop}%
\bibitem [{\citenamefont {Dobrev}\ \emph {et~al.}(1977)\citenamefont {Dobrev},
  \citenamefont {Mack}, \citenamefont {Petkova}, \citenamefont {Petrova},\ and\
  \citenamefont {Todorov}}]{Dobrev:1977qv}%
  \BibitemOpen
  \bibfield  {author} {\bibinfo {author} {\bibfnamefont {V.~K.}\ \bibnamefont
  {Dobrev}}, \bibinfo {author} {\bibfnamefont {G.}~\bibnamefont {Mack}},
  \bibinfo {author} {\bibfnamefont {V.~B.}\ \bibnamefont {Petkova}}, \bibinfo
  {author} {\bibfnamefont {S.~G.}\ \bibnamefont {Petrova}},\ and\ \bibinfo
  {author} {\bibfnamefont {I.~T.}\ \bibnamefont {Todorov}},\ }\href
  {https://doi.org/10.1007/BFb0009678} {\emph {\bibinfo {title} {{Harmonic
  Analysis on the n-Dimensional Lorentz Group and Its Application to Conformal
  Quantum Field Theory}}}},\ Vol.~\bibinfo {volume} {63}\ (\bibinfo {year}
  {1977})\BibitemShut {NoStop}%
\bibitem [{\citenamefont {Ort\'{i}n}(2002)}]{Ortin}%
  \BibitemOpen
  \bibfield  {author} {\bibinfo {author} {\bibfnamefont {T.}~\bibnamefont
  {Ort\'{i}n}},\ }\bibfield  {title} {\bibinfo {title} {A note on
  {L}ie-{L}orentz derivatives},\ }\href
  {https://doi.org/10.1088/0264-9381/19/15/101} {\bibfield  {journal} {\bibinfo
   {journal} {Classical and Quantum Gravity}\ }\textbf {\bibinfo {volume}
  {19}},\ \bibinfo {pages} {L143} (\bibinfo {year} {2002})}\BibitemShut
  {NoStop}%
  \bibitem [{\citenamefont {HiguchiLinearised}(1991)}]{HiguchiLinearised}%
  \BibitemOpen
  \bibfield  {author} {\bibinfo {author} {\bibfnamefont {A.}~\bibnamefont
  {Higuchi}},\ }\bibfield  {title} {\bibinfo {title} {{Linearized gravity in de Sitter spacetime as a representation of $SO(4,1)$}},\ }\href {https://doi.org/10.1088/0264-9381/8/11/011} {\bibfield
  {journal} {\bibinfo  {journal} {Classical and Quantum Gravity}\ }\textbf {\bibinfo
  {volume} {8}},\ \bibinfo {pages} {2005-2021} (\bibinfo {year} {1991})}\BibitemShut
  {NoStop}
  \bibitem [{\citenamefont {Higuchiforb}(1987)}]{Higuchiforb}%
  \BibitemOpen
  \bibfield  {author} {\bibinfo {author} {\bibfnamefont {A.}~\bibnamefont
  {Higuchi}},\ }\bibfield  {title} {\bibinfo {title} {{Forbidden mass range for spin-2 field theory in de Sitter spacetime}},\ }\href {https://doi.org/10.1016/0550-3213(87)90691-2} {\bibfield
  {journal} {\bibinfo  {journal} {Nuclear Physics B}\ }\textbf {\bibinfo
  {volume} {282}},\ \bibinfo {pages} {397-436} (\bibinfo {year} {1987})}\BibitemShut
  {NoStop}
\bibitem [{\citenamefont {Homma}\ and\ \citenamefont
  {Tomihisa}(2021{\natexlab{c}})}]{Homma}%
  \BibitemOpen
  \bibfield  {author} {\bibinfo {author} {\bibfnamefont {Y.}~\bibnamefont
  {Homma}}\ and\ \bibinfo {author} {\bibfnamefont {T.}~\bibnamefont
  {Tomihisa}},\ }\bibfield  {title} {\bibinfo {title} {{The spinor and tensor fields with higher spin on spaces of constant curvature}},\ }\href
  {https://doi.org/10.1007/s10455-021-09791-4} {\bibfield  {journal}
  {\bibinfo  {journal} {Annals of Global Analysis and Geometry}\ }\textbf {\bibinfo {volume} {60}},\
  \bibinfo {pages} {829-861} (\bibinfo {year} {2021}{\natexlab{c}})}\BibitemShut
  {NoStop}%
\bibitem [{\citenamefont {Hirai}(1962)}]{Hirai1}%
  \BibitemOpen
  \bibfield  {author} {\bibinfo {author} {\bibfnamefont {T.}~\bibnamefont
  {Hirai}},\ }\bibfield  {title} {\bibinfo {title} {On infinitesimal operators
  of irreducible representations of the lorentz group of $n$-th order},\ }\href
  {https://doi.org/10.3792/pja/1195523460} {\bibfield  {journal} {\bibinfo
  {journal} {Proc. Japan Acad.}\ }\textbf {\bibinfo {volume} {38}},\ \bibinfo
  {pages} {83} (\bibinfo {year} {1962})}\BibitemShut {NoStop}%
\bibitem [{\citenamefont {Hirai}(1965)}]{Hirai2}%
  \BibitemOpen
  \bibfield  {author} {\bibinfo {author} {\bibfnamefont {T.}~\bibnamefont
  {Hirai}},\ }\bibfield  {title} {\bibinfo {title} {The characters of
  irreducible representations of the lorentz group of $n$-th order},\ }\href
{https://doi.org/10.3792/pja/1195522333} {\bibfield  {journal} {\bibinfo
  {journal} {Proc. Japan Acad.}\ }\textbf {\bibinfo {volume} {41}},\ \bibinfo
  {pages} {526} (\bibinfo {year} {1965})}\BibitemShut {NoStop}%
\bibitem [{\citenamefont {Wong}(1974)}]{Wong}%
  \BibitemOpen
  \bibfield  {author} {\bibinfo {author} {\bibfnamefont {M.~K.~F.}\
  \bibnamefont {Wong}},\ }\bibfield  {title} {\bibinfo {title} {Unitary
  representations of ${SO}(n, 1)$},\ }\href {https://doi.org/10.1063/1.1666496}
  {\bibfield  {journal} {\bibinfo  {journal} {Journal of Mathematical Physics}\
  }\textbf {\bibinfo {volume} {15}},\ \bibinfo {pages} {25} (\bibinfo {year}
  {1974})},\ \Eprint {https://arxiv.org/abs/https://doi.org/10.1063/1.1666496}
{https://doi.org/10.1063/1.1666496} \BibitemShut {NoStop}%
   \bibitem [{\citenamefont {Sun}(2021)}]{Sun}%
   \BibitemOpen
  \bibfield  {author} {\bibinfo {author} {\bibfnamefont {Z.}\ \bibnamefont
  {Sun}},\ } \bibfield  {title} {\bibinfo {title} {{A note on the representations of SO(1,d+1),}}}\ (\bibinfo  {publisher} {arXiv},\ \bibinfo {year} {2021}),\ \Eprint {https://doi.org/10.48550/arXiv.2111.04591}{arXiv:2111.04591} \BibitemShut
{NoStop}%
 \bibitem [{\citenamefont {Gizem}(2021)}]{Gizem}%
   \BibitemOpen
  \bibfield  {author} {\bibinfo {author} {\bibfnamefont {G.}\ \bibnamefont
  {Seng{\"o}r}},\ } \bibfield  {title} {\bibinfo {title} {{The de Sitter group and its presence at the late-time boundary,}}}\ \href
  {https://arxiv.org/abs/2206.04719}{\bibfield  {journal} {\bibinfo
  {journal} {PoS}\ }\textbf {\bibinfo {volume} {2022}},\ \bibinfo
  {pages} {CORFU2021, 356} }\BibitemShut {NoStop}%
  \bibitem [{\citenamefont {Gazeau}(2022)}]{Gazeau}%
   \BibitemOpen
  \bibfield  {author} {\bibinfo {author} {\bibfnamefont {M.}\ \bibnamefont
  {Enayati}},\, \bibinfo {author} {\bibfnamefont {J.~P.}~\bibnamefont
  {Gazeau}}, \bibinfo {author} {\bibfnamefont {H.}~\bibnamefont
  {Pejhan}} and\ \bibinfo {author} {\bibfnamefont {A.}~\bibnamefont
  {Wang}},\ } \bibfield  {title} {\bibinfo {title} {{The de Sitter group and its representations: a window on the notion of de Sitterian elementary systems,}}}\ (\bibinfo  {publisher} {arXiv},\ \bibinfo {year} {2022}),\ \Eprint {https://arxiv.org/abs/2201.11457}{arXiv:2201.11457} \BibitemShut
{NoStop}
\bibitem [{\citenamefont {Dixmier}(1960)}]{Dixmier}%
  \BibitemOpen
  \bibfield  {author} {\bibinfo {author} {\bibfnamefont {J.}~\bibnamefont
  {Dixmier}},\ }\bibfield  {title} {\bibinfo {title} {Sur les représentations
  de certains groupes orthogonaux},\ }\href@noop {} {\bibfield  {journal}
  {\bibinfo  {journal} {Compt. Rend.}\ }\textbf {\bibinfo {volume} {250}},\
  \bibinfo {pages} {3263} (\bibinfo {year} {1960})}\BibitemShut {NoStop}%
  \bibitem [{\citenamefont {Kosmann}(1971)}]{Kosmann}%
  \BibitemOpen
  \bibfield  {author} {\bibinfo {author} {\bibfnamefont {Y.}~\bibnamefont
  {Kosmann}},\ }\bibfield  {title} {\bibinfo {title} {Dérivées de {L}ie des
  spineurs},\ }\href {https://doi.org/10.1007/BF02428822} {\bibfield  {journal}
  {\bibinfo  {journal} {Annali di Mat. Pura Appl. (IV)}\ }\textbf {\bibinfo
  {volume} {91}},\ \bibinfo {pages} {317} (\bibinfo {year} {1971})}\BibitemShut
  {NoStop}%
\bibitem [{\citenamefont {Todorov}(1978)}]{Todorov_1978}%
  \BibitemOpen
  \bibfield  {author} {\bibinfo {author} {\bibfnamefont {I.~T.}\ \bibnamefont
  {Todorov}}, \bibinfo {author} {\bibfnamefont {M.~C.}\ \bibnamefont {Mintchev}},\
  and\ \bibinfo {author} {\bibfnamefont {V.~B.}~\bibnamefont {Petkova}},\
  }\href {} {\emph
  {\bibinfo {title} {Conformal Invariance in Quantum Field Theory}}},\  (\bibinfo  {publisher} {Scuola Normale Superiore},\ \bibinfo {address}
  {Pisa},\ \bibinfo {year} {1978})\BibitemShut {NoStop}
\bibitem [{\citenamefont {Trautman}(1993)}]{Trautman_1993}%
  \BibitemOpen
  \bibfield  {author} {\bibinfo {author} {\bibfnamefont {A.\ \bibnamefont
  {Trautman}}},\
  }\href {} {\emph
  {\bibinfo {title} {Spin structures on hypersurfaces and the spectrum of the Dirac operator on spheres}}},\  (\bibinfo  {publisher} {in: Spinors, Twistors, Clifford Algebras and Quantum Deformations},\ \bibinfo {address}
  {Kluwer Academic Publishers},\ \bibinfo {year} {1993})\BibitemShut {NoStop}
  \bibitem [{\citenamefont {Letsios}(2023)}]{Letsios_in_progress}%
  \BibitemOpen
  \bibfield  {author} {\bibinfo {author} {\bibfnamefont {V. A.\ \bibnamefont
  {Letsios}}},\
  } \bibfield  {title} {\bibinfo {title} {{(Non-)unitarity of strictly and partially massless fermions on de Sitter space II: an explanation based on the group-theoretic properties of the spin-3/2 and spin-5/2 eigenmodes,}}}\
  {\bibfield  {journal} {\bibinfo  {journal} {In preparation}}}\ \BibitemShut {NoStop}
  \bibitem [{\citenamefont {Letsios_arxiv_long}(2022)}]{Letsios_arxiv_long}%
   \BibitemOpen
  \bibfield  {author} {\bibinfo {author} {\bibfnamefont {V. A.}\ \bibnamefont
  {Letsios}},\ } \bibfield  {title} {\bibinfo {title} {{The (partially) massless spin-3/2 and spin-5/2 fields in de Sitter spacetime as unitary and non-unitary representations of the de Sitter algebra,}}}\ (\bibinfo  {publisher} {arXiv},\ \bibinfo {year} {2022}),\ \Eprint {https://arxiv.org/abs/2206.09851}{arXiv:2206.09851v3} \BibitemShut{NoStop}
  %
\end{thebibliography}
\end{document}